\documentstyle[11pt,amssymb,epsf]{article}

\def\ben{\begin{equation}}
\def\een{\end{equation}}

\let\a=\alpha    
    
  \let\n=\nu

\let\C=\Chi

\def\nn{\nonumber} \def\bd{\begin{document}} \def\ed{\end{document}}
\def\ds{\documentstyle} \let\fr=\frac \let\bl=\bigl \let\br=\bigr
\let\Br=\Bigr \let\Bl=\Bigl
\let\bm=\bibitem
\let\na=\nabla
\let\pa=\partial \let\ov=\overline
\newcommand{\be}{\begin{equation}}
\newcommand{\ee}{\end{equation}}
\def\ba{\begin{array}}
\def\ea{\end{array}}
\def\ft#1#2{{\textstyle{{\scriptstyle #1}\over {\scriptstyle #2}}}}
\def\fft#1#2{{#1 \over #2}}
\def\del{\partial}
\def\vp{\varphi}
\def\sst#1{{\scriptscriptstyle #1}}
\def\oneone{\rlap 1\mkern4mu{\rm l}}
\def\td{\tilde}
\def\wtd{\widetilde}
\def\ie{\rm i.e.\ }
\def\dalemb#1#2{{\vbox{\hrule height .#2pt
        \hbox{\vrule width.#2pt height#1pt \kern#1pt
                \vrule width.#2pt}
        \hrule height.#2pt}}}
\def\square{\mathord{\dalemb{6.8}{7}\hbox{\hskip1pt}}}
\newcommand{\ho}[1]{$\, ^{#1}$}
\newcommand{\hoch}[1]{$\, ^{#1}$}
\newcommand{\bea}{\begin{eqnarray}}
\newcommand{\eea}{\end{eqnarray}}
\newcommand{\ra}{\rightarrow}
\newcommand{\lra}{\longrightarrow}
\newcommand{\Lra}{\Leftrightarrow}
\newcommand{\ap}{\alpha^\prime}
\newcommand{\bp}{\tilde \beta^\prime}
\newcommand{\tr}{{\rm tr} }
\newcommand{\Tr}{{\rm Tr} }
\def\0{{\sst{(0)}}}
\def\1{{\sst{(1)}}}
\def\2{{\sst{(2)}}}
\def\3{{\sst{(3)}}}
\def\4{{\sst{(4)}}}
\def\5{{\sst{(5)}}}
\def\6{{\sst{(6)}}}
\def\7{{\sst{(7)}}}
\def\8{{\sst{(8)}}}
\def\n{{\sst{(n)}}}
\def\cA{{{\cal A}}}
\def\cF{{{\cal F}}}
\def\tV{\widetilde V}
\def\tW{\widetilde W}
\def\tH{\widetilde H}
\def\tE{\widetilde E}
\def\tF{\widetilde F}
\def\tA{\widetilde A}
\def\im{{{\rm i}}}
\def\tY{{{\wtd Y}}}
\def\ep{{\epsilon}}
\def\vep{{\varepsilon}}
\def\R{\rlap{\rm I}\mkern3mu{\rm R}}
\def\bD{{{\bar D}}}

\def\R{\rlap{\rm I}\mkern3mu{\rm R}}
\def\bD{{{\bar D}}}
\def\R{{{\Bbb R}}}
\def\C{{{\Bbb C}}}
\def\H{{{\Bbb H}}}
\def\CP{{{\Bbb C}{\Bbb P}}}
\def\RP{{{\Bbb R}{\Bbb P}}}
\def\Z{{{\Bbb Z}}}
\def\bA{{{\Bbb A}}}
\def\bB{{{\Bbb B}}}
\def\bC{{{\Bbb C}}}
\def\bR{{{\Bbb R}}}
\def\bD{{{\Bbb D}}}
\def\bE{{{\Bbb E}}}
\def\bZ{{{\Bbb Z}}}
\def\Re{{{\frak{Re}}}}
\def\Im{{{\frak{Im}}}}
\def\cosec{{\,\hbox{cosec}\,}}
\def\Gm{{\Gamma_{\!\! -}}}
\def\Gp{{\Gamma_{\!\! +}}}
\def\stan{{standard }}
\def\nonstan{{supernumerary }}

\def\cosech{{\hbox{cosech}}}

\def\etcyc{{\hbox{and cyclic}}}

\newcommand{\tamphys}{\it Center for Theoretical Physics,
Texas A\&M University, College Station, TX 77843, USA}
\newcommand{\umich}{\it Michigan Center for Theoretical Physics,
University of Michigan\\ Ann Arbor, MI 48109, USA}
\newcommand{\upenn}{\it Department of Physics and Astronomy,
University of Pennsylvania\\ Philadelphia,  PA 19104, USA}
\newcommand{\SISSA}{\it  SISSA-ISAS and INFN, Sezione di Trieste\\
Via Beirut 2-4, I-34013, Trieste, Italy}

\newcommand{\newton}{\it Isaac Newton Institute for Mathematical Sciences,\\
20 Clarkson Road,  University of Cambridge,
Cambridge CB3 0EH, UK}

\newcommand{\ihp}{\it Institut Henri Poincar\'e\\
  11 rue Pierre et Marie Curie, F 75231 Paris Cedex 05}

\newcommand{\damtp}{\it DAMTP, Centre for Mathematical Sciences,
 Cambridge University\\ Wilberforce Road, Cambridge CB3 OWA, UK}
\newcommand{\itp}{\it Institute for Theoretical Physics, University of
California\\ Santa Barbara, CA 93106, USA}

\newcommand{\auth}{ 
G.W. Gibbons\hoch{\sharp},
Sean A. Hartnoll\hoch{\sharp} and C.N. Pope\hoch{\ddagger\flat}}

\begin{document}
\begin{flushright}
\hfill{DAMTP-2002-96}\ \ \ {CTP TAMU-18/02}\\
{August 2002}\ \ \
{hep-th/0208031}
\end{flushright}


\begin{center}
{ \Large {\bf Bohm and Einstein-Sasaki Metrics, Black Holes and
Cosmological Event Horizons}}

\vspace{5pt}
\auth

\vspace{3pt}
{\hoch{\sharp}\damtp}

\vspace{3pt}
{\hoch{\ddagger}\tamphys}

\vspace{3pt}
{\hoch{\flat}\newton}

\vspace{3pt}

\underline{ABSTRACT}
\end{center}

    We study physical applications of the Bohm metrics, which are
infinite sequences of inhomogeneous Einstein metrics on spheres and
products of spheres of dimension $5\le d \le 9$.  We prove that all
the Bohm metrics on $S^3\times S^2$ and $S^3\times S^3$ have negative
eigenvalue modes of the Lichnerowicz operator acting on transverse
traceless symmetric tensors, and by numerical methods we establish
that Bohm metrics on $S^5$ have negative eigenvalues too.  General
arguments suggest that all the Bohm metrics will have negative
Lichnerowicz modes.  These results imply that generalised
higher-dimensional black-hole spacetimes, in which the Bohm metric
replaces the usual round sphere metric, are classically unstable.  We
also show that the classical stability criterion for Freund-Rubin
solutions, which are products of Einstein metrics with anti-de Sitter
spacetimes, is the same in all dimensions as that for black-hole
stability, and hence such solutions based on the Bohm metrics will
also be unstable. We consider possible endpoints of the instabilities,
and in particular we show that all Einstein-Sasaki manifolds give
stable solutions. Next, we show how analytic continuation of Bohm
metrics gives Lorentzian metrics that provide counterexamples to a
strict form of the Cosmic Baldness conjecture, but they are
nevertheless consistent with the intuition behind the cosmic No-Hair
conjectures.  We indicate how these Lorentzian metrics may be created
``from nothing'' in a no-boundary setting. We argue that Lorentzian
Bohm metrics are unstable to decay to de Sitter spacetime.  Finally,
we argue that noncompact versions of the Bohm metrics have infinitely
many negative Lichernowicz modes, and we conjecture a general
relationship between Lichnerowicz eigenvalues and non-uniqueness of
the Dirichlet problem for Einstein's equations.

\pagebreak
\setcounter{page}{1}

\tableofcontents
\addtocontents{toc}{\protect\setcounter{tocdepth}{3}}
\vfill\eject

\section{Introduction}

   The properties of higher-dimensional black holes, and more
generally spacetimes with event horizons, have come to play an
increasingly important role in physics.  This is not only for the
purely theoretical reason that they may throw light on some hitherto
intractable problems of black holes in 3+1 spacetime dimensions, but
also because, if current ideas about large extra dimensions are
correct, then such black holes may possibly be created by high energy
collisions in accelerator experiments, and their behaviour might be
accessible to direct observation \cite{Giddings, Dimopoulos}.  As well
as having possible applications to laboratory scale physics, higher
dimensional black holes, and other types of horizons such as
cosmological event horizons \cite{GibbonsHawking1}, may also have
played an important role in the early universe.

   Many of the properties of black holes and event horizons in higher
dimensions are very similar to their counterparts in 3+1
dimensions. For example, the analogue of the spherically symmetric
Schwarzschild black hole exists in all dimensions, with the 2-sphere
of the four-dimensional solution replaced by a $(D-2)$ sphere in $D$
dimensions.  Likewise, there is an obvious higher-dimensional analogue
of the usual four-dimensional de Sitter spacetime. Moreover, subject
to the strict condition of asymptotic flatness, the former and their
charged versions are unique \cite{Ida1,Ida2}.  However, if one drops
the condition of strict asymptotic flatness, by allowing other compact
Einstein metrics in place of the usual round sphere in the spatial
sections at constant radius, then in higher dimensions there are many
more possibilities for black hole solutions, even on manifolds with the
same topology as the higher dimensional Schwarzschild solution.  This
is because of the remarkable fact, discovered by Bohm \cite{bohm},
that for $5 \le d\le 9$, the sphere $S^d$ carries infinitely many
other inhomogeneous Einstein metrics, in addition to its usual round
metric.  One might wonder whether the resulting black hole solutions
in spacetime dimensions $7\le D\le 11$ could arise during scattering
processes. This depends upon their stability. In this paper, we find
evidence, and in some cases proofs, that they are in fact unstable.
To do so we use methods developed in \cite{gh}, which showed that the
stability depends on the non-negativity of the spectrum of the
operator
\be
\Delta_{\rm stab} = \Delta _L +\fft{\Lambda}{d-1}\, \Big(
4 -\fft{(5-d)^2}{4}\Big)\,,\label{stabdef}
\ee
where $\Delta _L$ is the Lichnerowicz Laplacian on 
transverse traceless second rank symmetric tensor fields on the 
$d$-dimensional compact Einstein space, which satisfies $R_{ab} = \Lambda\, 
g_{ab}$.  We obtain numerical results establishing the existence of
negative Lichnerowicz modes in Bohm metrics on $S^5$, hence demonstrating
the instability of seven-dimensional black holes constructed using these
metrics.  We present general arguments suggesting that all the other
Bohm sphere metrics will have negative Lichnerowicz modes too.
 
   Bohm also showed the existence of infinitely many inhomogeneous
Einstein metrics on the products of spheres $S^{N_1}\times S^{N_2}$,
for $4\le N_1+N_2\le 9$, with $N_1\ge 2$ and $N_2 \ge 2$.  The lowest
dimensional such examples are on $S^3 \times S^2$.  We have also
investigated the associated topologically non-trivial black holes
(including the homogeneous product metric), and we prove that these are
unstable. It is also known that $S^3 \times S^2$ admits infinitely
many other homogeneous Einstein metrics, the so called $T^{p,q} \equiv
Spin(4)/U(1)$ spaces. Of these, only $T^{1,1}$ admits Killing spinors,
and in fact only the black hole associated with $T^{1,1}$ is stable
\cite{gh}. There are in addition some other {\it inhomogeneous}
metrics on $S^3 \times S^2$ with Killing spinors; these are examples
of Einstein-Sasaki metrics.  We show here that the associated black
holes using these metrics are stable.  One might think therefore
that an unstable black hole based on the usual product $S^3 \times
S^2$ metric would evolve dynamically into one based on one of the
$S^3\times S^2$ Einstein-Sasaki spaces.
Presumably however, in doing so the
area of the event horizon must increase.  Now at fixed ``mass
parameter'' we  show the areas of the horizons of the Einstein-Sasaki
horizons with topology $S^3 \times S^2$
 are less than those of the $S^3 \times S^2$ Bohm metrics. 
Thus in the evolution, the mass parameter  would have to increase,
which seems rather paradoxical.   
Another way to say this is that at fixed temperature (which is
proportional to the inverse of the mass parameter) the entropy
of the Einstein-Sasaki metrics is less than that of the Bohm metrics.

     In addition to constructing black holes, one may use Bohm's metrics
to obtain static inhomogeneous Lorentzian solutions of the Einstein
equations with a positive cosmological constant, which are
topologically the same as the static de Sitter metric.  Like de Sitter
spacetime, they contain a cosmological event horizon
\cite{GibbonsHawking1}.  As such they provide counterexamples in
$\le D\le 9$ spacetime dimensions to the long standing and hitherto
intractable Cosmic Baldness conjecture \cite{BoucherGibbons}, which
would be a generalisation of Israel's uniqueness theorem to cover the
de Sitter situation. The area of the cosmological event horizon in the
Lorentzian Bohm metrics is
smaller than that in de Sitter spacetime.  We believe therefore that
they are unstable, and that under a small perturbation they would
evolve, at least within the event horizon of a given observer, to a
static de Sitter-like state. If this is the case, then although
evading the strict letter of the Cosmic Baldness conjecture they would
respect the spirit of the weaker No-Hair conjecture,.  The latter
asserts that apart from unstable cases of measure zero, the generic
solution should settle down to a de Sitter-like state within the
horizon of any given observer.  This is all that is needed to justify
the usual intuition behind inflationary models of the early universe.
Another possible application for these generalised de Sitter spacetimes
would be as tunnelling metrics.

    The plan of this paper is as follows.  In section 2 we review the
link established in \cite{gh} between black-hole stability and the
spectrum of the Lichnerowicz Laplacian on the compact $d$-dimensional
Einstein space $M_d$ that forms the constant-radius spatial sections.
We also show that this stability criterion is identical, for all
dimensions of $M_d$, to that for the stability of Freund-Rubin type 
AdS$_n\times M_d$ solutions of gravity coupled to a $d$-form.  We 
then discuss a lower bound on the spectrum of the Lichnerowicz
operator, based on considerations of the Weyl curvature, that was 
considered in \cite{pagepope,pagepope2}.

   In section 3 we give a detailed discussion of the Bohm metrics, and
we exhibit negative modes of the Lichnerowicz operator in some of these
backgrounds.  In certain cases, including all the Bohm metrics on
$S^3\times S^2$ and on $S^3\times S^3$, we obtain an analytic proof of
the existence of negative modes.  Intuition leads one to expect
negative modes in all the Bohm metrics, and we back this up with some
numerical results in certain examples where an analytic proof is
lacking.  There are also non-compact examples of Bohm metrics, which
are Ricci-flat.  We give analytic proofs that the non-compact Bohm
metrics on $\R^3 \times S^2$ and $\R^3\times S^3$, recently considered
by Kol \cite{Kol}, have negative modes of the Lichnerowicz Laplacian.
Again, intuition leads one to expect negative modes for all the
non-compact Bohm examples.

   In section 4 we discuss Einstein-Sasaki metrics, which may be defined
as odd-dimensional Einstein metrics $ds^2$ whose cone $d\hat s^2 = dr^2 
+ r^2\, ds^2$ is Ricci-flat and K\"ahler.  They admit Killing
spinors, and we use this fact to obtain a lower bound on the spectrum
of the Lichnerowicz operator.  In particular we use this to demonstrate 
that the associated black
holes are always stable.  Likewise, this establishes that Freund-Rubin
compactifications using Einstein-Sasaki manifolds will always be stable. 

    Section 5 contains a description of the analytic
continuation of the Bohm metrics to give Lorentzian spacetimes that
are generalisations of de Sitter spacetime.  These 
provide counterexamples to the Cosmic Baldness conjecture.  
In section 6 we try to relate the existence of negative modes for
the Lichnerowicz Laplacian to the non-uniqueness of the Dirichlet
problem for the Einstein equations.  Section 7 gives our main
conclusions, and points to some other applications of Bohm metrics.
For example, they can provide magnetic monopole solutions in Kaluza-Klein
theory. An appendix gives further details about our numerical
techniques, and includes some graphs illustrating the behaviour of the 
metric functions in the Bohm solutions.

\section{Stability and the Lichnerowicz Laplacian}

	Our first aim will be to study the classical stability of two
types of spacetime constructed using a general positive curvature
Einstein metric $M_d$.  The first of these comprises generalisations
to higher dimensions of the four-dimensional Schwarzschild black hole,
in which the spatial 2-sphere at constant radius is generalised to a
higher-dimensional Einstein space $M_d$.  The second class of examples
comprises Freund-Rubin type solutions AdS$_n\times M_d$ to a theory of
Einstein gravity coupled to a $d$-form field strength.  As we shall
show below, the classical stability criteria for both of these classes
of spacetimes are expressible as the {\it same} criterion on the
spectrum of the Lichnerowicz operator acting on transverse traceless
symmetric 2-index tensors on $M_d$.  To set the stage for this
discussion, we begin in section \ref{stabcritsec} with a general
discussion of the Lichnerowicz operator, reviewing the manner in which
it arises from a consideration of the second variation of the
Einstein-Hilbert action.  We then review the black-hole stability
\cite{gh} and AdS$_n\times M_d$ stability \cite{dnp,dfghm} criteria in
sections \ref{blacksec} and \ref{adssec}, and in section
\ref{weyltensorsec} we review an argument given in
\cite{pagepope,pagepope2} which shows how a lower bound on the
Lichnerowicz spectrum can be obtained by considering the eigenvalues
of the Weyl tensor.

\subsection{Stability criteria}\label{stabcritsec}

We begin by considering the Einstein-Hilbert action in $d$ dimensions
\be
S=\int_M \sqrt{g}\, d^dx\, (R- (d-2)\, \Lambda)\,,\label{ehact}
\ee
whose Euler-Lagrange equations give the Einstein equation
\be
R_{ab} = \Lambda \, g_{ab} \,.
\ee
Under the perturbation
\be
g_{ab} \to g_{ab} + h_{ab} \,,
\ee
one finds that up to quadratic order in $h_{ab}$, the action $S$ is
given on-shell by $S=S_0 + S_1 + S_2 +\cdots$, with
\bea
&&S_0=2\Lambda\, \int_M\sqrt{g}\, d^dx\,,\qquad S_1=0\,,\nn\\
&&S_2= \int_M\sqrt{g}\, d^dx\, (-\ft14 h^{ab}\, \Delta_2\, h_{ab} 
   + \ft14 h\, \Delta_0\, h + \ft12 (\nabla_a\, h^{ab})^2)\,,
\eea
where $h\equiv h_a{}^a$, $\square \equiv \nabla^a\, \nabla_a$ and
\be
\Delta_0\, h \equiv -\square h + \ft12(d-2)\, \Lambda\, h\,,\qquad
\Delta_2\equiv \Delta_L -2\Lambda\,.\label{del2lich}
\ee
Here $\Delta_L$ is the Lichnerowicz Laplacian operator acting on 
symmetric rank two tensors,
\be
\label{eq:lich}
\Delta_L\, h_{ab} \equiv - \square \,h_{ab} -2 R_{acbd}\,  h^{cd} 
+ R_{ca}\,  h^c{}_b + R_{cb} \, h^c{}_a\,. 
\ee
If we consider a transverse traceless perturbation,
\be
\nabla^a h_{a b} = 0\,, \qquad h^a{}_a = 0 \,,
\ee
then the second variation of the action is simply given by
\be
S_2 = -\ft14 \int_M\sqrt{g}\, d^dx\, h^{ab}\, \Delta_2\, h_{ab}\,.
\label{quadratic}
\ee

   In a perturbative classical stability analysis one asks whether there
are finite energy solutions to the linearised equations of motion
\be
\Delta_2\, h_{ab}=0\label{eq:perteqns}
\ee
that grow exponentially in time. This question has
been studied in some depth recently for generalised vacuum black hole
spacetimes in $D=d+2$ dimensions \cite{gh} and also for spacetimes
that are a direct product of $D-d$ dimensional anti-de Sitter with a
$d$ dimensional compact Einstein manifold \cite{dfghm}. The anti-de
Sitter spacetimes are supported by a gauge field of appropriate rank.

\subsection{Generalised Schwarzschild-Tangherlini spacetimes}
\label{blacksec}

Generalised Schwarzschild-Tangherlini black holes have the form
\be
d\hat s^2 = - \left[1-\left(\frac{\ell}{r}\right)^{d-1} \right] dt^2 +
\frac{dr^2}{\left[1-\left(\frac{\ell}{r}\right)^{d-1} \right]} 
+ r^2 \, ds^2_d ,
\ee
where $\ell$ is a constant and $ds^2_d$ is the metric on a
$d$ dimensional compact Einstein manifold $B$ with the curvature
normalised to be that of $S^d$
\be\label{eq:curv}
R_{\alpha\beta}=(d-1) g_{\alpha\beta} .
\ee
The black hole solution has vanishing cosmological constant.

   It was found in \cite{gh} that the dangerous mode for
instability is a transverse tracefree eigenfunction of the
Lichnerowicz Laplacian on the Einstein manifold $B$.
\bea
\label{eq:mode}
&&\hat h_{0 a} = \hat h_{1 a} = 0 , \nonumber \\
&&\hat h_{\alpha\beta} = h_{\alpha\beta}(x)\,  r^2 \, \phi(r) 
\, e^{\im\, \omega t} ,
\eea
where $x$ are coordinates on $B$ and
\be
\label{eq:eval}
\Delta_L\, h_{\alpha\beta} = \lambda h_{\alpha\beta} .
\ee
Here $\Delta_L$ is the Lichnerowicz Laplacian on $B$. The
stability of the spacetime was found to depend on the spectrum
$\{\lambda\}$ of the Lichnerowicz Laplacian acting on
transverse tracefree modes on $B$. Concretely, if the spectrum
contains an eigenvalue that is too negative, the spacetime is
unstable:
\be
\label{eq:crit}
\lambda_{min} < \lambda_c \equiv 4 - \frac{(5-d)^2}{4} 
\quad\Leftrightarrow\quad \mbox{instability} .
\ee
This result follows from considering the behaviour of the radial
dependence of the perturbation (\ref{eq:mode}), $\phi(r)$. When the
criterion (\ref{eq:crit}) is satisfied, the solution for $\phi(r)$
that decays at infinity also oscillates logarithmically in the
interior. This allows it to be matched to a solution that is well
behaved at the horizon $r=\alpha$. Thus a finite energy mode exists in
this case and the spacetime is unstable.

   It is perhaps worth pointing out here that  a form of Birkhoff
theorem holds for the metrics we are considering. In other words
if we had assumed a general time dependent metric of the form
\be
ds^2 = - e^{-2\phi(r,t)} \,dt^2 + e^{2 \psi (r,t)} \,  dr^2 +
Y^2(r,t)\,  ds^2_d\,,
\ee
where $ds^2_d$ is a $d$-dimensional {\sl time-independent} Einstein
metric with scalar curvature $d(d-1)$, we would have found, on
imposing the Einstein equations,  that the metrics had to  be
static. In fact this result also holds if the metric is coupled to
a 2-form field strength  (in the electric case) or a  $d$-form
field strength the magnetic case. It also holds if one includes a
cosmological term. It means that when perturbing the static metric
we must consider time dependent perturbations of the transverse or
base  Einstein metric $ds^2_d$. This is another way of seeing why we need
information about the the spectrum of the Lichnerowicz operator on
this space.

   We shall not give a detailed proof of Birkhoff's theorem here, but
merely indicate how to modify an existing treatment of Wiltshire
\cite{wiltshire} which assumes $SO(d+1)$-invariance, i.e that $ds^2_d$
is the unit round metric on $S^d$. Wiltshire gives a proof which also
covers the case when Gauss-Bonnet terms are present. The argument he
presents will not go over to the case of a general Einstein metric,
since it makes special use of properties of its Riemann tensor. Thus
in what follows we ignore that term, which means we set $\tilde \alpha
=0$ in his equations. It is an interesting question to ask whether
Birkhoff's theorem remains true when one includes a Gauss-Bonnet term.

  The discussion depends upon whether $\partial _\alpha Y$ is
spacelike, timelike or null. We assume the first case, and make a
coordinate choice such that $Y=r$.  The field equation $R_t{}^r =0$
(equation (6b) in his paper)  then yields
$$
\partial_t \psi=0.
$$
The equation $R_t{}^t+ R_r{}^r =0$ then gives
$$
\partial_r \phi + \partial_r \psi =0\,.
$$
This means that $\phi + \psi =f(t)$, where $f(t)$ is an
arbitrary differentiable function of $t$. By choice of the
coordinate $t$,  $f(t)$ may   be taken   to vanish. It follows
that both $\phi$ and $\psi$ are independent of time $t$, and the
metric is therefore static. The remaining field equations show that it takes
the Schwarzschild-Tangherlini  form.

\subsection{Anti-de Sitter product spacetimes}\label{adssec}

These are solutions to a system with a $d$-form field strength
minimally coupled to gravity
\bea
d\hat s^2 = ds^2_{AdS_{D-d}} + ds^2_d , \nonumber \\
F_d = \left(\frac{2(D-2)(d-1)}{D-d-1}\right)^{1/2} \mbox{vol}_B
\eea
where we have taken the same normalisation for the curvature of $B$ as
previously (\ref{eq:curv}).

   It was shown in \cite{dnp,dfghm} that the dangerous mode is a
transverse tracefree mode on the manifold $B$ multiplying a scalar on
the AdS spacetime
\bea
&&\hat h_{0 a} = \hat h_{1 a} = 0 , \nonumber \\
&&\hat h_{\alpha\beta} = h_{\alpha\beta}(x) \, \phi(y) ,
\eea
where $y$ are the coordinates on the AdS. The mode
$h_{\alpha\beta}(x)$ on $B$ is an eigentensor as
in (\ref{eq:eval}). From the AdS point of view, $\phi(y)$ is seen as a
scalar field with mass given by \cite{dfghm}
\be
\label{eq:mass}
m^2 = \lambda -2(d-1) .
\ee
Instability of massive scalars on AdS spacetime is expressed in terms
of the Breitenlohner-Freedman bound \cite{bf,mt}. In our units, this
reads
\be
\label{eq:BF}
m^2 \left(\frac{D-d-1}{d-1}\right)^2 < - \frac{(D-d-1)^2}{4} 
\quad\Leftrightarrow\quad \mbox{instability} .
\ee
Using the value of the mass in (\ref{eq:mass}), the criterion
(\ref{eq:BF}) is just
\be
\lambda_{min} < \lambda_c \equiv 4 - \frac{(5-d)^2}{4} 
\quad\Leftrightarrow\quad \mbox{instability} \,.\label{instability}
\ee
This is immediately seen to be the same as the criterion found for the
black hole spacetimes (\ref{eq:crit}). This is an intriguing match.

\subsection{Lichnerowicz Laplacian and the Weyl tensor}\label{weyltensorsec}

    In order to make estimates of the lowest eigenvalue of the 
Lichnerowicz Laplacian, it is convenient first to rewrite (\ref{eq:lich}) in
terms of the Weyl tensor.  Since we are assuming that the metric is
Einstein, with $R_{ab}=\Lambda\, g_{ab}$, this is given in $d$ dimensions 
by
\be
C_{abcd} = R_{abcd} - \fft{\Lambda}{d-1}\, (g_{ac}\, g_{bd} - g_{ad}\, 
            g_{bc})\,.
\ee
Thus the Lichnerowicz Laplacian becomes
\be
\Delta_L\, h_{ab} = -\square \, h_{ab} - 2 C_{acbd}\, h^{cd}
    + \fft{2d\, \Lambda}{d-1}\, h_{ab}\,.\label{lich2}
\ee

   A method for obtaining a lower bound on the smallest eigenvalue of
$\Delta_L$ was introduced in \cite{pagepope}.  One considers the integral
of $3(\nabla_{(a}\, h_{bc)})^2$, which, after performing an integration
by parts and using the transversality and tracelessness of $h_{ab}$, gives
\bea
\int_M h^{ab}\, \Delta_L\, h_{ab} &=& 
  \int_M[3(\nabla_{(a}\, h_{bc)})^2 -4 h_{ab}\, R_{abcd}\, h^{cd} + 
     4\Lambda\, h_{ab}\, h^{ab}]\,,\nn\\
&=&\int_M[3(\nabla_{(a}\, h_{bc)})^2 -4 h_{ab}\, C_{abcd}\, h^{cd} + 
     \fft{4d\, \Lambda}{d-1}\, h_{ab}\, h^{ab}]\,\nn\\
&\ge & \int_M( -4 h_{ab}\, C_{abcd}\, h^{cd} + 
     \fft{4d\, \Lambda}{d-1}\, h_{ab}\, h^{ab})\,.\label{intbound}
\eea
Viewing $C_{abcd}$ as a map acting on symmetric traceless tensors $h_{ab}$,
we can define its eigenvalues $\kappa$ by
\be
C_{acbd}\, h^{cd} = \kappa\, h_{ab}\,.
\ee

   For homogeneous spaces, we therefore have a simple inequality
\be
\Delta_L \ge  \fft{4d\, \Lambda}{d-1} -4\kappa_{\rm max}\,,\label{hombound}
\ee
where $\kappa_{\rm max}$ is the largest eigenvalue of the Weyl
tensor.  Equality is attained if the corresponding eigentensor
$h_{ab}$ is a Staeckel tensor, satisfying $\nabla_{(a}\, h_{bc)}=0$.
The bound (\ref{hombound}) was derived for homogeneous Einstein
7-manifolds $M_7$ in \cite{pagepope}.  It was also shown that in the
case of $U(1)$ bundles over the Einstein-K\"ahler product 6-manifolds
$S^2\times \CP^2$ and $S^2\times S^2\times S^2$, the eigentensor that
maximises $\lambda$ is in fact Staeckel, and thus the equality in
(\ref{hombound}) is attained \cite{pagepope,pagepope2}.

   For inhomogeneous spaces, such as the Bohm metrics that we shall be
studying later in this paper, the bound (\ref{intbound}) must be kept in 
its integrated form, and so we have
\be
\int_M h^{ab}\, \Delta_L\, h_{ab}\ge 
\int_M  \Big(\fft{4d\, \Lambda}{d-1}- 
4\kappa_{\rm max}(x)\Big)\, h_{ab}\, h^{ab}
\,,
\ee
where $\kappa_{\rm max}(x)$ represents the (position dependent)
largest eigenvalue of the Weyl tensor.  If we define $\kappa_{\rm
max}^0$ to be the largest value attained by any of the eigenvalues of
the Weyl tensor anywhere in $M$, $\kappa_{\rm max}^0\equiv
\hbox{sup}_x\, \kappa_{\rm max}(x)$, then we obtain the inequality
\be
\Delta_L \ge  \fft{4d\, \Lambda}{d-1} -4\kappa^0_{\rm max}\,.
\label{inhombound}
\ee
It is clear, however, that this bound is not likely to be very sharp, 
especially if $\kappa_{\rm max}$ depends strongly on position.

   We can, nevertheless, extract a general feature of the spectrum of
the Lichnerowicz operator from the above considerations, namely that
in Einstein spaces of positive Ricci tensor, the minimum eigenvalue
tends to be lowered by having a large positive eigenvalue of the Weyl
tensor. Indeed, we can see from (\ref{inhombound}) that if the largest
Weyl tensor eigenvalue is not sufficiently positive, then $\Delta_L$
could never be negative or zero.  We shall see when we study the Bohm
metrics in detail that in these cases the Weyl tensor can in fact
have sufficiently large eigenvalues that the Lichnerowicz spectrum
includes negative eigenvalues.  By using a Rayleigh-Ritz variational 
method we shall be able to obtain upper bounds on the lowest eigenvalue
of the Lichnerowicz Laplacian, allowing us in some cases to give
an analytic proof of the existence of negative-eigenvalue modes.

\section{The Bohm Einstein Metrics on $S^N$ and 
$S^{N_1}\times S^{N_2}$}

\subsection{Description of the Bohm construction}

    Bohm's construction \cite{bohm} gives rise to a countable infinity
of Einstein metrics with positive Ricci tensor on the spheres $S^N$
for $5\le N\le 9$, and on the product topologies $S^{N_1}\times
S^{N_2}$ for $5\le N_1+N_2\le 9$ with $N_1\ge 2$ and $N_2\ge 2$. We
can use these metrics in the black-hole and AdS$\times M_d$ spacetimes
of the previous section.

   The starting point for the construction is the following ansatz for
metrics of cohomogeneity one,
\be
ds^2 = dt^2 + a^2\, d\Omega_p^2 + b^2\, d \wtd\Omega_q^2\,,
\label{bohmans}
\ee
where $a$ and $b$ are functions of the radial variable $t$, and
$d\Omega_p^2$ and $d\wtd\Omega_q^2$ are the standard metrics on the
unit spheres $S^p$ and $S^q$.  An elementary calculation shows that in
the orthonormal frame $e^0=dt$, $e^i =a\, \bar e^i$, $\hat e^{\a} =
b\, \bar e^{\a}$, where $\bar e^i\, \bar e^i = d\Omega_p^2$ and
$\bar e^{\a}\, \bar e^{\a}=d\wtd\Omega_q^2$, the components of the Riemann
tensor are given by
\bea
&& R_{0i0j} = -\fft{\ddot a}{a}\, \delta_{ij}\,,\qquad
  R_{0 \a 0\beta} = -\fft{\ddot b}{b}\, \delta_{\a\beta}\,,\qquad
  R_{i\a j\beta} = -\fft{\dot a\, \dot b}{a\, b}\, \delta_{ij}\,
\delta_{\a\beta}\,,\nn\\
&&R_{ijk\ell} =  \fft{1-\dot a^2}{a^2}\, (
\delta_{ik}\, \delta_{k\ell} -\delta_{i\ell}\, \delta_{jk})\,,\qquad
R_{\a\beta\gamma\delta} =
\fft{1-\dot b^2}{b^2}\, (
\delta_{\a\gamma}\, \delta_{\beta\delta} -\delta_{\a\delta}\,
\delta_{\beta\gamma})\,,\label{bohmriem}
\eea
The components of the Ricci tensor are given by
\bea
R_{00} &=& -\fft{p\, \ddot a}{a} - \fft{q\, \ddot b}{b}\,,\nn\\
R_{ij} &=& - \Big[ \fft{\ddot a}{a} + \fft{q\,
\dot a\, \dot b}{a\, b} + \fft{(p-1)\, (\dot a^2-1)}{a^2}\Big]\,
\delta_{ij}\,,\nn\\
R_{\a\beta} &=& -\Big[ \fft{\ddot b}{b} + \fft{p\,
\dot a\, \dot b}{a\, b} + \fft{(q-1)\, (\dot b^2-1)}{b^2}\Big]\,
\delta_{\a\beta}\,.\label{ricci}
\eea
    
   The Einstein equations $R_{ab}=\Lambda\, g_{ab}$ give rise
to two second-order differential equations for $a$ and $b$,
\bea 
\fft{\ddot a}{a} + \fft{q\,
\dot a\, \dot b}{a\, b} + \fft{(p-1)\,(\dot a^2-1)}{a^2}
&=&-\Lambda\,,\nn\\
\fft{\ddot b}{b} + \fft{p\,
\dot a\, \dot b}{a\, b} + \fft{(q-1)\,(\dot b^2-1)}{b^2}
&=& -\Lambda\,,\label{einstein}
\eea
together with the first-order constraint
\be
\fft{p\, (p-1)\, (\dot a^2-1)}{a^2} + 
\fft{q\, (q-1)\, (\dot b^2-1)}{b^2} + \fft{2p\, q\, \dot a\, \dot b}{a\, b}
+ (p+q-1)\, \Lambda=0\,.\label{constraint}
\ee
We shall adopt the conventional normalisation, when considering Einstein
metrics with positive $\Lambda$, of taking
\be
\Lambda= p+q\,,
\ee
which is one less than the total dimension of the space.

   When $p>1$ and $q>1$, which we shall be considering here, the general
solution of the Einstein equations is not known explicitly.  A well-known
special solution is 
\be
a=\sin t\,,\qquad b=\cos t\,,\label{bohm0}
\ee
in which case the metric (\ref{bohmans}) becomes just the standard 
round metric on $S^{p+q+1}$,
\be
ds^2 = dt^2 + \sin^2t\, d\Omega_p^2 + \cos^2t\, d\wtd\Omega_q^2\,,
\ee
written as a foliation by $S^p\times S^q$.  This can easily be
recognised as the metric on the unit $S^{p+q+1}$ by introducing
coordinates $x^A$ on $\R^{p+q+2}$, subject to the unit-radius
constraint $x^A\, x^A=1$, and then introducing orthogonal unit-vectors
$m^A$ and $n^A$ in $\R^{p+q+2}$, such that a general point on the unit
$S^{p+q+1}$ in $\R^{p+q+2}$ can be written as
\be
x^A= m^A\, \sin t + n^A\, \cos t\,.
\ee

   A second well-known special solution to the Einstein equations is
\be
a= \sqrt{\fft{p}{\Lambda}}\, \sin\Big( \sqrt{\fft{\Lambda}{p}}\, t\Big) \,,
\qquad b = \sqrt{\fft{q-1}{\Lambda}}\,,\label{bohm1}
\ee
with, using our conventional choice, $\Lambda=p+q$.
This gives the standard homogeneous Einstein metric on $S^{p+1}\times S^q$,
\be
ds^2 = dt^2 + \fft{p}{\Lambda}\, 
\sin^2\Big( \sqrt{\fft{\Lambda}{p}}\, t\Big) \, d\Omega_p^2 + 
\fft{q-1}{\Lambda}\, d\wtd\Omega_q^2\,.
\ee
There is an analogous solution for $S^p\times S^{q+1}$ too.  Since
there is obviously always a discrete transformation under which the
roles of the spheres $S^p$ and $S^q$ are interchanged, we shall not in
general bother to mention the symmetry-related possibility.

    It is shown in \cite{bohm} that the Einstein equations
(\ref{einstein}) and (\ref{constraint}) admit a countably infinite
number of solutions giving rise to inequivalent metrics that extend
smoothly onto manifold of topology $S^{p+q+1}$, and another countable
infinity of solutions for which the metrics extend smoothly onto
manifolds of topology $S^{p+1}\times S^q$.  We shall denote these
metrics by Bohm$(p,q)_n$, where the integer $n$ runs over $n=0,2,4,\ldots$
for the $S^{p+q+1}$ sequence, and $n=1,3,5\ldots$ for the $S^{p+1}\times S^q$
sequence.  The standard unit metric (\ref{bohm0}) on $S^{p+q+1}$ corresponds
to Bohm$(p,q)_0$, and the standard product Einstein metric (\ref{bohm1})
on $S^{p+1}\times S^q$ corresponds to Bohm$(p,q)_1$.

    The higher metrics Bohm$(p,q)_n$ with $n\ge 2$ are all inhomogeneous.
The radial coordinate $t$ runs between endpoints which can be taken to be
0 and $t_f$, defined by the vanishing of one or other of the metric
functions $a$ and $b$.   The metric extends onto the corresponding degenerate
orbit because the associated metric function vanishes like $t$ or 
$(t_f-t)$, so that one has a regular collapsing of $p$-spheres 
or $q$-spheres like in the origin of spherical polar coordinates.  
For all the $S^{p+q+1}$ metrics Bohm$(p,q)_{2m}$ one has
\bea
&&a(0)=0\,,\qquad \dot a(0)=1\,,\qquad b(0)=b_0\,,\qquad \dot b(0)=0\,;\nn\\
&&
a(t_f) = a_0\,,\qquad \dot a(t_f)=0\,,\qquad 
b(t_f)= 0\,,\qquad \dot b(t_f)= -1\,.\label{bohmeven}
\eea
On the other hand, for the $S^{p+1}\times S^q$ metrics Bohm$(p,q)_{2m+1}$
one has
\bea
&&a(0)=0\,,\qquad \dot a(0)=1\,,\qquad b(0)=b_0\,,\qquad \dot b(0)=0\,;
\nn\\
&&a(t_f) = 0\,,\qquad \dot a(t_f)= -1\,,\qquad 
b(t_f)= \td b_0\,,\qquad \dot b(t_f)=0\,.\label{bohmodd}
\eea
The functions $a$ and $b$ are strictly positive for $0<t<t_f$, and the
quantities $a_0$, $b_0$ and $\td b_0$ are certain constants.

   Plots of the metric functions $a$ and $b$ for various Bohm metrics
are presented in the Appendix.  These have been obtained by performing a
numerical integration of the Einstein equations (\ref{einstein}).  It can be
seen that as the index $n$ for Bohm$(p,q)_n$ increases, the metrics 
rapidly become approximations to the ``double-cone'' Einstein metric
\be
ds^2 = dt^2 + \fft{1}{(p+q-1)}\,\, \sin^2 t\, [ (p-1)\, d\Omega_p^2
                + (q-1)\, d\wtd\Omega_q^2]\,,\label{bohmcone}
\ee
for most of the range of the radial coordinate.
The metric (\ref{bohmcone}) itself is singular at the apexes $t=0$ and
$t=\pi$, since near to each of these points one has a collapse of
$S^p\times S^q$ surfaces.  The actual Bohm$(p,q)_n$  metrics with large
$n$ deviate from (\ref{bohmcone}) just in the vicinity of the apexes,
instead approaching the forms given in (\ref{bohmeven}) or (\ref{bohmodd}).
It is interesting to note that (\ref{bohmcone}) is in fact the singular limit
both of the regular $S^{p+q+1}$ sequence Bohm$(p,q)_{2m}$ and the 
regular $S^{p+1}\times S^q$ sequence Bohm$(p,q)_{2m+1}$.

   Note that in the case of the Bohm metrics Bohm$(p,q)_{2m+1}$ with
the topology $S^{p+1}\times S^q$, the fact that the metric function
$b(t)$ never vanishes means that we can replace the associated round
sphere $S^q$ with its metric $d\wtd \Omega_q^2$ in (\ref{bohmans}) by
{\it any} Einstein space $Q_q$ of dimension $q$, whose (positive)
Ricci curvature is normalised to $\wtd R_{\a\beta}=(q-1)\, \td
g_{\a\beta}$, and we will again have a complete and non-singular 
Bohmian metric in $(p+q+1)$ dimensions, now with the topology 
$S^{p+1}\times Q_q$.  The Einstein space $Q_q$ could itself be 
taken to be a Bohm metric such as Bohm$(2,2)_n$, Bohm$(2,3)_n$
or Bohm$(3,2)_n$.

\subsection{Estimates and bounds for Lichnerowicz in Bohm metrics}

\subsubsection{Eigenvalues of the Weyl tensor}\label{weylevs}

   We saw earlier, in section \ref{weyltensorsec}, that positive eigenvalues
of the Weyl tensor tend to drive the lowest mode of the Lichnerowicz 
more negative.  Accordingly, we can gain insights into the bounds on the
spectrum of the Lichnerowicz operator in the Bohm metrics by studying 
the Weyl tensor.  From (\ref{bohmriem}), and our choice of normalisation 
where $R_{ab}=(p+q)\, g_{ab}$, we have
\bea
&&C_{0i0j}=x_1\, \delta_{ij}\,,\quad C_{0\a0\beta}= x_2\, \delta_{\a\beta}
\,,\quad C_{i\a j\beta}= x_3\, \delta_{ij}\, \delta_{\a\beta}\,,\nn\\
&&C_{ijk\ell} = x_4\, (\delta_{ik}\, \delta_{j\ell} -
\delta_{i\ell}\, \delta_{jk})\,,\quad
C_{\a\beta\gamma\delta} = x_5\, (\delta_{\a\gamma}\, \delta_{\beta \delta}
-\delta_{\a\delta}\, \delta_{\beta\gamma})\,,
\eea
where
\be
x_1=-1-\fft{\ddot a}{a}\,,\quad x_2 = -1-\fft{\ddot b}{b}\,,\quad
x_3= -1 -\fft{\dot a\, \dot b}{a\, b}\,,\quad
x_4= \fft{1-{\dot a}^2 -a^2}{a^2}\,,\quad
x_5 = \fft{1 - {\dot b}^2 -b^2}{b^2}\,.
\ee
Note that these coefficients $x_i$ are not all independent, and thus
\be
p\, x_1=-q\, x_2 = \ft12q\, (q-1)\, x_5 - \ft12 p\, (p-1)\, x_4\,,\quad
x_3= -\fft{p\, (p-1)\, x_4 + q\, (q-1)\, x_5}{2p\, q}\,.
\ee

   It is straightforward to see that with multiplicities ${\bf m}$, the 
traceless eigenvectors $h_{AB}$ and eigenvalues $\kappa$ of the Weyl
tensor are given by 
\bea
h_{0i}:&& {\bf m}=p\,,\qquad \kappa= -x_1\,,\nn\\
h_{0\a}:&& {\bf m}=q\,,\qquad \kappa= -x_2\,,\nn\\
\{h_{ij}\ |\  h_{ii}=0\}:&& {\bf m} =\ft12p\, (p+1)-1\,,
\qquad \kappa= -x_4\,,\nn\\
\{h_{\a\beta}\ |\ h_{\a\a}=0\}:&& {\bf m} =\ft12q\, (q+1)-1\,,\qquad 
\kappa= -x_5\,,\nn\\
h_{i\a} :&& {\bf m}=p\, q\,,\qquad \kappa =-x_3\,,
\eea
together with two eigenvectors of the form
\be
h_{00}= -p\, u -q\, v\,,\qquad h_{ij} = u\, \delta_{ij}\,,\qquad 
 h_{\a\beta} = v\, \delta_{\a\beta}\,,\label{balloonbohm}
\ee
for which the eigenvalues are given by the roots of a quadratic equation,
\bea
\kappa_{\pm}&=& \ft12(p-1)\, x_4 + \ft12 (q-1)\, x_5 \nn\\
&&\pm \fft{\sqrt{
p\, q\, [(p-1)\, x_4 + (q-1)\, x_5]^2 + (p+q+1)\, [p\, (p-1)\, x_4 
  -q\, (q-1)\, x_5]^2}}{2\sqrt{p\, q}}\,.\label{kappapm}
\eea
The coefficients $u$ and $v$ are then given by
\be
u = -2q\, \kappa_\pm -p\, (p-1)\, x_4 + q\, (q-1)\, x_5\,,\quad
v = 2p\, \kappa_\pm -p\, (p-1)\, x_4 + q\, (q-1)\, x_5\,.
\ee
In total, we have the expected $\ft12(p+q+1)(p+q+2) -1$ symmetric traceless 
eigenmodes in $(p+q+1)$ dimensions.

   Using the output of the numerical integration of the Einstein
equations for the Bohm metrics, we find that the eigenvalue of the
Weyl-tensor that achieves the largest positive value is $\kappa_+$
given by (\ref{kappapm}).  It is therefore in this sector that one can
expect to find the lowest-lying eigenmodes of the Lichnerowicz
operator.  We can see from (\ref{balloonbohm}) that the associated
eigenvector is of a type that may be thought of as a ``ballooning
mode.''  That is to say, it is a mode where one of the spheres $S^p$ or
$S^q$ tends to inflate at the expense of the other.  This accords with 
one's intuition, which would suggest that the most likely instability
for metrics with direct-product orbits would be ballooning modes of this
general type.

   Some examples of our numerical results for the maximum value of
the largest eigenvalue of the Weyl tensor are as follows.  For the 
Bohm$(2,2)_n$ metrics on $S^5$ and $S^3\times S^2$ we find from
(\ref{kappapm}) that $\kappa_+$ attains its maximum value at the endpoints
of the radial coordinate range, and so
\be
\kappa_{\rm max}^0 = \fft{5(1-b_0^2)}{3 b_0^2}\,.
\ee
In fact, as $n$ increases, the function $\kappa_+$ peaks more and more
strongly around the endpoints. For $n=0,\ldots,6$, we have approximately
\be
\kappa_{\rm max}^0=\{0,5,24.26, 118.45, 579.76, 3013.72,15106.9\}\,.
\ee
(The results for $n=0$ and $n=1$ are exact, since these the standard
homogeneous metrics on $S^5$ and $S^3\times S^2$.)  Using
(\ref{inhombound}), we obtain the lower bound
\be
\Delta_L \ge - \fft{20(1-4 b_0^2)}{3b_0^2}
\ee
on the spectrum of the Lichnerowicz operator on Bohm$(2,2)_n$.  From 
our numerical results for the values of $b_0$ for the first few examples, 
we find
\bea
\hbox{Bohm}(2,2)_0:&& \Delta_L \ge 20\,,\nn\\
\hbox{Bohm}(2,2)_1:&& \Delta_L \ge 0\,,\nn\\
\hbox{Bohm}(2,2)_2:&& \Delta_L \ge -77.04\,,\nn\\
\hbox{Bohm}(2,2)_3:&& \Delta_L \ge -453.8\,,\nn\\
\hbox{Bohm}(2,2)_4:&& \Delta_L \ge  -2342\,,\nn\\
\hbox{Bohm}(2,2)_5:&& \Delta_L \ge   - 11972\,,\nn\\
\hbox{Bohm}(2,2)_6:&& \Delta_L \ge  - 60407\,.\label{lowerbounds}
\eea
The bounds for $n=0$ and $n=1$ are in fact exactly attained,
corresponding to the the cases of the homogeneous $S^5$ and $S^3\times
S^2$ metrics respectively.  The zero-mode in the latter case is the
ballooning mode on $S^3\times S^2$.  As we mentioned previously, the
lower bounds we obtain for the inhomogeneous Bohm metrics are not 
expected to be very sharp.

   In general for the Bohm$(p,q)_n$ metrics we find
\be
\kappa^0_{\rm max} = \fft{(q-1)\, (p+q+1)\,(1-b_0^2)}{(p+1)\, b_0^2}\,,
\ee
and hence we have the lower bound
\be
\Delta_L \ge - \fft{4(p+q+1)\, [q-1-(p+q)\, b_0^2]}{(p+1)\, b_0^2}\,,
\ee
where as usual we have normalised the scale so that $R_{ab}=(p+q)\ g_{ab}$.

\subsubsection{Transverse tracefree ballooning modes}

  In order to study transverse tracefree perturbations, we consider a
metric of the form 
\be
ds^2 = c^2\, dt^2 + a^2 \, d\Omega_p^2 + b^2\, d\wtd\Omega_q^2\,.
\label{eq:pertmetric}
\ee
This is similar to (\ref{bohmans}), except that we have, for convenience,
introduce the coordinate gauge function $c(t)$ in the metric.
Substituting into the Einstein-Hilbert action
\be
S = \int \sqrt{g}\, d^dx\, \left[R - (d-2) (d-1) \right]\,,
\ee
(where $d=p+q+1$)
and omitting a constant factor equal to the volume of the product metric
on the unit $S^p\times S^q$, this gives 
\bea
S &=& \int a^p \, b^q\, c\,  \Big[ 2pq \frac{\dot{a} \dot{b}}{a b c^2} 
+ p(p-1) \frac{\dot{a}^2}{a^2 c^2} +
q(q-1) \frac{\dot{b}^2}{b^2 c^2} + \frac{p(p-1)}{a^2} + 
\frac{q(q-1)}{b^2}\nn\\
&&\qquad\qquad - (p+q)(p+q-1) \Big] \,dt \,.\label{sint}
\eea

    Ballooning modes in product metrics, in which one factor contracts
and the other expands, are Lichnerowicz zero modes and are typically
associated with instabilities \cite{dnp,pagepope,dfghm,gh}.  It is
reasonable to expect, therefore, as we argued in section \ref{weylevs},
that if instabilities were to arise in the Bohm metrics, they would be
associated with modes of a similar type.  We are therefore led to seek
a generalisation of ballooning modes to the warped product of spheres
present in (\ref{eq:pertmetric}). The perturbation
\be
a \to a\, \sqrt{1+ u}\,, \quad
b \to b\, \sqrt{1+ v} \,,\quad 
c = 1 \to \sqrt{1+\gamma} \,,\label{albega}
\ee
is tracefree at the linearised level if $\gamma+ p\, u + q\, v=0$, and
transverse if 
\be
\dot \gamma + (p\, \fft{\dot a}{a} + q\, \fft{\dot b}{b})\, \gamma
   - p\, \fft{\dot a}{a}\, u - q\, \fft{\dot b}{b}\, v=0\,.
\ee
These two conditions can be used in order to solve for $u$ and $v$ 
in terms of $\gamma$:
\be
\label{eq:tt}
u = \frac{\dot{\gamma}+ \left[ (q+1) \, \dot{b}\, b^{-1} + 
p \, \dot{a}\, a^{-1}\right]\, \gamma }
{p \, (\dot{a}\, a^{-1} - \dot{b}\, b^{-1})} \,,\quad 
v = \frac{\dot{\gamma}+ \left[ (p+1) \,\dot{a}\, a^{-1} + 
q \, \dot{b}\, b^{-1}\right]\,\gamma }
{q \, (\dot{b}\, b^{-1} - \dot{a}\, a^{-1})} \,.
\ee

   One is free to choose the function $\gamma$, which
completely determines the perturbation through (\ref{eq:tt}). However,
the Bohm$(p,q)_n$ metric has $n$ interior points at which
$(\dot{a}\, a^{-1} - \dot{b}\, b^{-1})$ vanishes, and hence the
expressions for $u$ and $v$ are singular for generic choices
of $\gamma$. This problem can be solved by inverting (\ref{eq:tt})
to give $\gamma$ in terms of $u$,
\be
\gamma = \frac{1}{a^p\,  b^{q+1}} \, \int a^p \, b^{q+1} \, 
\left[\frac{\dot{a}}{a} 
- \frac{\dot{b}}{b} \right] \, u \, dt \,.
\ee
The remaining function, $v$, is given by the tracefree condition,
$p\, u+q\, v+\gamma=0$.  One is now free to choose a
nonsingular function $u$ to obtain a perturbation that will at
worst be singular at the endpoints $t=0$, $t=t_f$. These
singularities can be avoided as described in the next
subsection. One simple choice is
\be
u = \frac{p-m}{p\,  a^m \, b^{p+q+1-m}}\,, \quad
v = \frac{-(p+1-m)}{q \, a^m\,  b^{p+q+1-m}}\,, \quad
\gamma = \frac{1}{a^m \, b^{p+q+1-m}}\,.\label{gamma}
\ee
We will use these expressions below in order to exhibit negative
Lichnerowicz modes in some of the Bohm metrics.

\subsubsection{Rayleigh-Ritz estimates for the lowest Lichnerowicz 
eigenvalue}

    In the eigenvalue problem $\Delta\, \phi= \lambda\, \phi$ for a 
self-adjoint operator $\Delta$, one can obtain an upper bound on the lowest 
eigenvalue by the Rayleigh-Ritz method, namely
\be
\lambda_{\rm min} \le \fft{\int_M \psi\, \Delta\, \psi}{\int_M \psi^2}\,,
\ee
with equality being achieved if the trial function $\psi$ is actually the
eigenfunction corresponding to the lowest eigenvalue.  In this section,
we shall apply this method to obtain an upper bound on the lowest
eigenvalue of the Lichnerowicz operator on TT modes in the Bohm metrics, 
and in particular, we shall find that there is a negative-eigenvalue
mode in some of cases we examine.

   The easiest cases to consider are the Bohm metrics
Bohm$(p,q)_{2m+1}$ on the product topologies $S^{p+1}\times S^q$.  In
these metrics, the function $b$ in (\ref{bohmans}) is nowhere
vanishing, and so we can take our trial function to be given by
(\ref{gamma}) with $m=0$:
\be
u = \frac{\ep}{b^{p+q+1}}\,, \quad
v = \frac{-(p+1)\,\ep}{q \, b^{p+q+1}}\,, \quad
\gamma = \frac{\ep}{b^{p+q+1}}\,.\label{gamma0}
\ee
Here we have introduced $\ep$ as a small constant order-parameter.  As
we shall see below, this trial function allows us to prove that 
certain of the Bohm metrics on products of spheres have negative eigenvalue
modes of the Lichnerowicz operator.

\subsection{Negative Lichnerowicz eigenvalues in Bohm metrics}

\subsubsection{Analytic results for negative modes for 
Bohm metrics on $S^3\times S^2$}

   It turns out that the easiest cases to study are the Bohm metrics
Bohm$(2,2)_{2m+1}$, whose topology is $S^3\times S^2$.  We are able to
obtain completely analytic and explicit results that prove the
existence of negative modes of the Lichnerowicz operator for all these
examples (for $m\ge 1$), and so we shall present the details for these
metrics here.  For these examples, we take $\gamma=\ep/b^5$ as our
trial function, as suggested by (\ref{gamma0}), implying that we have
$u= \ep/b^5$ and $v=-3\ep/(2b^5)$.  From the expansion of $S$ given in
(\ref{sint}), with the perturbation (\ref{albega}), we can easily
extract the terms quadratic in $\ep$, and so by comparing with
(\ref{del2lich}) and (\ref{quadratic}) we obtain a Rayleigh-Ritz bound
for the lowest eigenvalue $\lambda_{\rm min}$ of the Lichnerowicz
operator:
\be
\lambda_{\rm min} \le \fft{\int_0^{t_f} P\, dt}{\int_0^{t_f} Q\, dt}\,,
\label{rrfun}
\ee
where 
\bea
P &=& -\fft{3 a^2}{b^{10}} + \fft{7}{b^8} - \fft{30 a^2}{b^8}
  + \fft{7 {\dot a}^2}{b^8} - \fft{112 a\, \dot a\, \dot b}{b^9}
   - \fft{91 a^2\, {\dot b^2}}{2b^8}\,,\nn\\
Q &=& \fft{15 a^2}{2 b^8}\,,\label{pq22}
\eea
and $t_f=2 t_c$ is the upper limit of the range of the radial coordinate
$t$.  Note that in fact it suffices to evaluate the integrals 
only up to the mid-point 
$t=t_c$ in these cases, since the metric functions are symmetric about
$t=t_c$ here.

    Using the constraint (\ref{constraint}), with $p=q=2$, we can eliminate
the term involving ${\dot a}^2$ in $P$.  We also note that upon use of the
second-order equations  (\ref{einstein}), we can prove the identities
\bea
\fft{d}{dt}\, \Big(\fft{a\, \dot a}{b^8}\Big) &=& - 
\fft{10 a\, \dot a\, \dot b}{b^9} + \fft{1-4a^2}{b^8}\,,
\nn\\
\fft{d}{dt}\,\Big( \fft{a^2\, \dot b}{b^9}\Big) &=& 
 - \fft{10 a^2\, {\dot b}^2}{b^{10}} + \fft{a^2\, (1-4b^2)}{b^{10}}\,.
\eea
Since $a\, \dot a/b^8$ and $a^2\, \dot b/b^9$ vanish at both endpoints
of the full integration range, we can use
these in order to perform integrations by parts in the evaluation of
$\int P$.  Specifically, we use the former to remove the term in $P$
involving $\dot a\, \dot b$, and then using the latter, we find that
\be
\int_0^{t_f} P\, dt = -\fft{25}{2}\, \int_0^{t_f} \fft{a^2\, 
{\dot b}^2}{b^{10}}\, dt\,,\label{p22}
\ee
which is manifestly non-positive.  We therefore have the Rayleigh-Ritz bound
\be
\lambda_{\rm min} \le -\fft{5}{3}\, 
\fft{\int_0^{t_f} a^2\, b^{-10}\, {\dot b}^2\, dt}{
\int_0^{t_f} a^2\, b^{-8}\, dt}
\ee
for the lowest eigenvalue of the Lichnerowicz operator.
(Recall that we are working in units where $R_{ab}=4g_{ab}$.)
This proves that the Einstein metrics Bohm$(2,2)_{2m+1}$ have a
negative eigenvalue for the Lichnerowicz operator on transverse
traceless symmetric 2-index tensors, for $m\ge 1$.  (The case $m=0$ is
the standard product Einstein metric on $S^3\times S^2$, with
$b=\ft12$.  In this case the numerator gives zero, and in fact we exactly
saturate the upper bound, finding the known lowest eigenvalue $\Delta_L=0$ 
on the product metric.)

  Using the numerical results described in the Appendix, we find 
the following upper bounds on the lowest Lichnerowicz eigenvalue for
the Bohm$(2,2)_3$ and Bohm$(2,2)_5$ Einstein metrics on $S^3\times S^2$:
\bea
\hbox{Bohm}(2,2)_3:&& \lambda_{\rm min} \le -7.2766\,,\nn\\
\hbox{Bohm}(2,2)_5:&& \lambda_{\rm min} \le -198.008\,.\label{upperbounds1}
\eea
(Recall that we are normalising the metrics so that $R_{ab}=4\,
g_{ab}$.)  As one goes to higher examples Bohm$(2,2)_{2m+1}$ with
increasing $m$, one finds that the upper bound on the lowest
Lichnerowicz eigenvalue becomes increasingly negative, tending to
$-\infty$ in the limit as $m\longrightarrow \infty$.  Note that our
upper bounds (\ref{upperbounds1}) are considerably larger than the
rather crude lower bounds (\ref{lowerbounds}) that we obtained by
considering the eigenvalues of the Weyl tensor.

\subsubsection{Analytic results for negative modes for 
Bohm metrics on $S^3\times S^3$}

    For general values of $p$ and $q$, the analogous expression 
for the integrands $Y$ in (\ref{p22}) and $Q$ in (\ref{pq22}) that
appear in the numerator and denominator of the Rayleigh-Ritz functional 
(\ref{rrfun}) turn out to be
\bea
P&=& q^{-1}\, (p+q+1)^2\, (p^2-2p-3+p\, q -q)\, a^p\, b^{-2p-q-4}\, 
{\dot b}^2\,,\nn\\
Q&=& q^{-1}\, (p+q+1)\, (p+1)\, a^p\, b^{-2p-q-2}\,,\label{pqgen}
\eea
if we take the trial function $\gamma=b^{-p-q-1}$. In general, this
gives us a rather weak positive upper bound on the lowest Lichnerowicz
eigenvalue for the Bohm$(p,q)_{2m+1}$ metrics on $S^{p+1}\times S^q$.
In fact only for $p=q=2$, which we discussed above, and $p=2$, $q=3$,
does one get a non-positive bound from this choice of trial function.
Interestingly, for $p=2$, $q=3$ the numerator integrand $P$ in
(\ref{pqgen}) vanishes identically, and so we obtain the bound
\be
\lambda_{\rm min} \le0 \label{bound23}
\ee
in this case.  It is straightforward to see that for $m\ge1$ the 
trial function $\gamma=1/b^6$ does not give an eigenfunction, and hence
the inequality in (\ref{bound23}) is not saturated.  Thus we have
an analytic proof that for the Bohm$(2,3)_{2m+1}$ Einstein metrics on
$S^3\times S^3$ with $m\ge1$, the lowest eigenvalue of the Lichnerowicz
operator on TT symmetric tensors is strictly negative.

   For all other cases aside from Bohm$(2,2)_{2m+1}$ and
Bohm$(2,3)_{2m+1}$, the trial function $\gamma=b^{-p-q-1}$ does not
give a negative upper bound on $\lambda_{\rm min}$.  We believe that
this is a consequence of a non-optimal choice of trial function,
since the qualitative arguments would suggest the existence of
negative Lichnerowicz modes for all the Bohm metrics.  

\subsubsection{Numerical results for negative modes for Bohm$(2,2)_{2m}$ 
 metrics   }

   The analytic methods that allowed us to prove the existence of
negative modes of the Lichnerowicz operator in the $S^3\times S^2$ and
$S^3\times S^3$ Bohm metrics do not directly extend to any of the Bohm
metrics on $S^{p+q+1}$. The reason for this is that whilst our trial
function $\gamma=1/b^{p+q+1}$ is regular everywhere in the metrics on
$S^{p+1}\times S^q$,  it diverges at the right-hand endpoint of the 
range of the radial coordinate $t$ in the metrics on $S^{p+q+1}$.
A natural modification to the trial function to take account of this
is to interpolate smoothly between $\gamma=1/b^{p+q+1}$ on the left-hand side 
and $\gamma=1/a^{p+q+1}$ on the right-hand side of the range of $t$.
We have carried out this procedure numerically in the case of examples 
of the Bohm$(2,2)_{2m}$ metrics on $S^5$, and we find that indeed there
are negative modes of the Lichnerowicz operator, in accordance with
the qualitative arguments.  Specifically, we find approximately
\bea
\hbox{Bohm}(2,2)_2:&& \lambda_{\rm min} \le - 0.7937\,,\nn\\
\hbox{Bohm}(2,2)_4:&& \lambda_{\rm min} \le - 38.86\,,\nn\\
\hbox{Bohm}(2,2)_6:&& \lambda_{\rm min} \le - 1040.6\,.
\eea
These upper bounds are again all considerably larger than the corresponding
lower bounds in (\ref{lowerbounds}) that we obtained from the eigenvalues
of the Weyl tensor.

\subsection{Non-compact Bohm metrics}\label{noncompsec}

    A class of complete and non-singular non-compact metrics was also
constructed by Bohm \cite{Bohm2}.  These include examples where
the metric ansatz is again taken to be (\ref{bohmans}), but now the
metric is required to be Ricci-flat. These metrics have
been considered recently in \cite{Kol} in studies of the possibility
of topology change. It was shown in \cite{Bohm2} that
regular metrics exist in which $a(t)$ and $b(t)$ satisfy the boundary 
conditions
\be
a(0)=0\,,\qquad \dot a(0)=1\,,\qquad b(0)= b_0\,,\qquad \dot b(0)=0\,,
\ee
Unlike the previous compact examples, here regularity imposes no 
constraint on the allowed values for the constant $b_0$, and in fact
the value of $b_0$ now merely sets the overall scale of the metric.
Note that $b$ is everywhere non-vanishing, and so the there is an 
$S^q$ bolt at $t=0$.  The metrics are asymptotically conical, 
approaching cones over the standard product Einstein metric on $S^p\times
S^q$.

   A representative example is presented in the Appendix, for the case
of $p=2$, $q=2$.  

   The Rayleigh-Ritz method that we described earlier for finding an
upper bound on the smallest Lichnerowicz eigenvalue can be applied in
these non-compact Bohm metrics too.  In fact the trial function
$\gamma=b^{-p-q-1}$ can be considered here too, since it remains finite
everywhere and it falls off rapidly at large $t$.  We find that the 
numerator and denominator integrands are then again given by (\ref{pqgen}),
and so again we obtain a negative upper bound on the lowest Lichnerowicz
eigenvalue for the cases $p=q=2$, and $p=2$, $q=3$.  

    Evaluating the integrands numerically for the case $p=q=2$, 
we find that 
\be
\lambda_{\rm min} \le - \fft{0.110433}{b_0^2}\,.
\ee
For $p=2$, $q=3$, we find, as for Bohm$(2,3)_{2m+1}$, that the bound
is $\lambda_{\rm min}\le 0$, and we can again argue that since the
trial function $\gamma=1/b^6$ does not give an eigenfunction, we must
have $\lambda_{\rm min}<0$.  For all other $p$ and $q$, our choice of
trial function does not give a negative upper bound on the lowest
eigenvalue of the Lichnerowicz operator.  Again, we believe that this
is because the trial function is non-optimal in these other cases, since
general arguments suggest that the non-compact Bohm metrics should all have
negative eigenvalue modes of the Lichnerowicz operator.

\section{Einstein-Sasaki manifolds}

\subsection{Introduction and definition} 

    In this section we remind the reader that as well as the infinite
sequence of cohomogeneity one Bohm metrics that have featured in our
discussion, the manifold $S^3 \times S^2$ admits many other Einstein
metrics.  For example, it has been known for some time that there are
infinitely-many homogeneous but non-supersymmetric $T^{p,q}$ spaces,
corresponding to $U(1)$ bundles over $S^2\times S^2$ in which the
$U(1)$ fibres wind $p$ times over one $S^2$, and $q$ times over the
other. These all have the topology $S^3\times S^2$ and they all 
admit an Einstein metric.  Only $T^{1,1}$ admits Killing spinors.

    There are, by contrast, also many inequivalent supersymmetric
examples of $S^3\times S^2$ Einstein metrics, which do admit Killing
spinors.  They can thus be used in the AdS/CFT correspondence,
replacing $S^5$ in the D3 brane metric and its near-horizon limit.

   An Einstein-Sasaki metric may be defined as a $(2m+1)$-dimensional 
Einstein metric such that the cone over it is a Calabi-Yau metric
\be
ds^2 _{\rm Calabi-Yau} = d R^2 + R^2\,  ds^2_{\rm Einstein-Sasaki}\,,
\label{cyes}
\ee
or in other words, the cone is a Ricci-flat K\"ahler metric.
The Killing spinors in the Einstein-Sasaki metric come by direct 
projection from the 
covariantly-constant  spinors of the Calabi-Yau metric.
If one uses the complex structure $J$  of the Calabi-Yau to act on the
Euler vector of the cone, $ R\, { \partial \over \partial R}$,
one gets a Killing vector on the Einstein-Sasaki manifold
with constant magnitude, and thus we may write locally
\be
ds^2 _{\rm Einstein-Sasaki}= (d \psi +A) ^2 + 
    ds^2_{\rm Einstein-K\ddot{a}hler}\,,
\ee
where $J{ \partial \over \partial R}= {\partial \over \partial \psi}$
and $ ds^2_{\rm Einstein-K\ddot{a}hler}$ is locally Einstein-K\"ahler
with positive scalar curvature.
Globally the $U(1)$ action generated by ${\partial \over \partial \psi}$
may be free (in which case one speaks of a regular Sasaki
structure) and the base is a smooth Einstein-K\"ahler manifold,
or it may have fixed points in which case the Einstein-K\"ahler base
has orbifold singularities. The total space however will still be smooth.
We give a more detailed discussion of the relation between the 
Einstein-K\"ahler and Einstein-Sasaki spaces below.

   Taking $\CP^2$ or $\CP^1 \times \CP^1$ as the Einstein-K\"ahler
base metric gives the standard homogeneous Sasaski metrics on $S^5$ or
$T^{1,1}$ respectively. If the fibration is regular the only remaining
possible base metrics for 5-dimensional Einstein-Sasaki metrics are
inhomogeneous metrics on del Pezzo surfaces, \ie $\CP^2$ blown up
at $k$ points, with $3\le k\le 8$, giving Einstein-Sasaki metrics on the
connected sum of $k$ copies of $S^3 \times S^2$. \footnote {It is
worth remarking that {\sl any} 5-dimensional closed simply-connected
spin manifold with no torsion in the second homology group is
diffeomorphic to a connected sum of copies of $S^3 \times S^2$}.

   Recently Boyer, Galicki et al. \cite{Galicki1, Galicki2, Galicki3,Galicki4}
have constructed many inhomogeneous Einstein-Sasaki $(2n-1)$ metrics on
the links $L_f =C_f \cap S^{2n+1} $ of weighted homogeneous
polynomials $f$ on $\C^{n+1}$.  The notation is as follows: $C_f \in
\C^{n+1}$ is the zero set $f(z)=0$ of the polynomial, and $S^{2n+1}$
is the standard sphere. One readily sees that the Hopf fibration
descends to $L_f$, and this gives the fibration associated to the
Sasaki structure.  Note that this description is purely topological.
The metric is obtained indirectly by means of an existence proof.  The
present state of the art is that there are at least 14 inequivalent
Einstein-Sasaki structures on $S^3 \times S^2$ \cite{Galicki4}.  Of these, only
$T^{1,1}$ is homogeneous, and so if used in the AdS/CFT correspondence
the other 13 examples would give supersymmetric vacua with no
R-symmetry.

The volume of $T^{1,1}$ is well known to be
$16 \pi ^3 \over 27$. 
According to \cite{Herzog}, the volume of $L_f$
is given, in five dimensions,  by
\be
 {\pi ^3 \over 27w} (|w|-d)^3,
\ee 
where $d$ is the degree of $f$, ${\bf w}=(w_0,w_1,w_2,w_3)$ are the
weights, $|{\bf w}|=w_0+w_1+w_2+w_3$ and $w=w_0w_1w_2w_3$. The two
inhomogeneous Einstein-Sasaki metrics on $S^3 \times S^2$ constructed
in \cite{Galicki1}  are both of
degree 256 and have weights ${\bf w}=(11,49,69,128)$ and ${\bf
w}=(13,35,81,128)$. They therefore have volumes $\pi^3
\over 27 \times 4760448$ and $\pi ^3 \over27 \times 4717440$
respectively.  These may be compared with the volume of the product
metric on $S^3 \times S^2$, which is $ \pi ^3 \over \sqrt{2}$, and of
the limiting singular double cone Bohm metric, which is $ 2\pi^3 \over
3$.
  
\subsection{Einstein-Sasaki manifolds as $U(1)$ bundles over 
Einstein-K\"ahler manifolds}

    In this subsection, we present what is essentially a review of how
Einstein-Sasaki manifolds can be constructed as $U(1)$ bundles over
Einstein-K\"ahler manifolds, focusing in particular on the
construction of the Killing spinors.  The construction can be applied
to obtain Einstein-Sasaki manifolds in any odd dimension, and so we
shall give the construction for this general case.

   Suppose we have an Einstein-K\"ahler metric $g_{ab}$ on a manifold
$M_n$ of (even) dimension $n=2m$.  By the standard formulae of Kaluza-Klein
reduction, the $(n+1)$-dimensional metric
\be
d\hat s^2 = (d\psi + A)^2 + ds^2
\ee
has Ricci tensor $\hat R_{AB}$ whose frame components are given by
\be
\hat R_{ab} = R_{ab} -\ft12 F_a{}^c\, F_{bc}\,,\quad
\hat R_{00}= \ft14 F_{ab}\, F^{ab}\,,\quad 
\hat R_{0a}= \ft12 \nabla^b\, F_{ab}
\ee
where $F=dA$, and $\hat e^0=d\psi+A$, $\hat e^a=e^a$.  Taking $F_{ab}=
\mu\, J_{ab}$, where $J_{ab}$ is the K\"ahler form on $M_n$, we therefore
have
\be
\hat R_{ab} = (\Lambda-\ft12 \mu^2)\, \hat g_{ab}\,,\quad
\hat R_{00} = \ft14 n\, \mu^2\,,\quad \hat R_{0a}=0\,,
\ee
where $R_{ab}=\Lambda\, g_{ab}$ in $M_n$, and so the $(n+1)$-dimensional 
metric $d\hat s^2$ will be Einstein, $\hat R_{AB}=\hat\Lambda\, \hat g_{AB}$, 
and
\be
\hat \Lambda= n\,,\qquad \Lambda= n+2\,,
\ee
provided that we take $\mu=2$.

   The covariant exterior derivative on spinors, $\hat D\equiv d + \ft14
\hat \omega_{AB}\, \Gamma^{AB}$, is easily seen to be given by
\be
\hat D= D -\ft14 A\, J_{ab}\, \Gamma^{ab} + \ft12 J_{ab}\, e^b\, \Gamma^{0a}
       - \ft14 J_{ab}\, d\psi\, \Gamma^{ab}\,,\label{spincov}
\ee
where $D=d+\ft14\omega_{ab}\, \Gamma^{ab}$ is the covariant exterior
derivative on spinors in the base space $M_n$.  Note that since 
$n$ is necessarily even, the spinors in the total space $\hat M$ have the
same dimension as those in base space $M_n$, and so we do not need 
to make any tensor-product decomposition of the Dirac matrices. 
The equation for Killing
spinors in the $(n+1)$-dimensional bundle space $\hat M$, in the 
normalisation $\hat R_{AB}= n\, \hat g_{AB}$ that we established above,
is simply $\hat D_A\, \eta = \ft12\, \im\, \sigma\, \Gamma_A\, \eta$,
where $\sigma=\pm1$.  From (\ref{spincov}), this gives the equations
\bea
D_a\, \eta -A_a\, \fft{\del\eta}{\del\psi} &=& \ft12 J_{ab}\, 
\Gamma^{0b}\, \eta + \ft12\, \im\, \sigma\, \Gamma_a\, \eta\,,\nn\\
\fft{\del\eta}{\del\psi}&=& \ft14 J_{ab}\, \Gamma^{ab}\, \eta +\ft12\im\, 
\sigma\, \Gamma_0\, \eta\,.\label{spinred}
\eea

    As is well known, the Einstein-K\"ahler space $M_n$ admits a 
gauge-covariantly constant spinor $\varepsilon$, satisfying 
\be
{\cal D}_a\, \varepsilon - \im\, e\, A_a\, \varepsilon=0\,,
\ee
where as above we have $dA=F=2J$, and $e$ is the electric charge
carried by $\varepsilon$.  This can be determined by examining the
integrability condition $[{\cal D}_a,{\cal D}_b]\, \varepsilon = \ft14
R_{abcd}\, \Gamma^{cd}\, \varepsilon -2\im\, e\, J_{ab}\,
\varepsilon$.  Multiplying by $\Gamma^{ab}$, this gives $\im\, n\,
(n+2)\,\varepsilon = 4e\, J_{ab}\, \Gamma^{ab}\,\varepsilon$.  It is a
straightforward exercise to calculate the eigenvalues of the matrix $J_{ab}\,
\Gamma^{ab}$, and to show, in particular, that in general it has only 
two singlet eigenvalues, which are $\pm\im\, n$.  It is these singlets that
are associated with the gauge-covariantly-constant spinor $\varepsilon$
(and its charge conjugate), and so we can deduce that 
\be
e = \ft14 (n+2)\,.
\ee
One can also then easily show that $\Gamma^0\, \varepsilon = \sigma\, 
\varepsilon$, where $\sigma=\pm1$.  From the second equation in 
(\ref{spinred}) we therefore deduce that if we take $\eta=f(\psi)\, 
\varepsilon$
we shall have
\be
f = e^{\fft14 (n+2)\, \im\, \psi}\,\,,
\ee
and then the first equation in (\ref{spinred}) confirms that indeed
$\varepsilon$ satisfies
\be
D_a\, \varepsilon  -\ft14 \im\, (n+2)\, A_a\, \varepsilon=0\,.
\ee
In other words, we have proved that if $\varepsilon$ is the
gauge-covariantly constant spinor in the Einstein-K\"ahler manifold
$M_n$, then $ \eta=e^{\fft14 (n+2)\,\im\,  \psi}\, \varepsilon$ is a Killing
spinor in the $U(1)$ bundle over $M_n$, which is therefore an
Einstein-Sasaki manifold $\hat M$. The conjugate spinor satisfies the
Killing-spinor equation with the opposite sign on the right-hand side.
Lifted up further using (\ref{cyes}), one obtains the conjugate pair
of covariantly-constant spinors in the Ricci-flat K\"ahler cone 
over the Einstein-Sasaki manifold.  The situation is summarised in the
table below.

\bigskip\bigskip
\centerline{
\begin{tabular}{|c|c|c|}\hline
$2m$-Dimensional  &  $(2m+1)$-Dimensional   & $(2m+2)$-Dimensional \\
Einstein-K\"ahler &  Einstein-Sasaki        & Calabi-Yau Cone      \\
                  &                         &                      \\
 $\Lambda>0$      &  $\Lambda >0$           & $R_{ab}=0$           \\
                  &                         &                      \\
$(D_a -\im\, e\, A_a)\, \varepsilon=0$ &
                     $D_a\, \eta = \pm \im\, m\, \Gamma_a\, \eta$ &
                                             $D_a\, \eta =0$ \\ \hline
\end{tabular}}
\bigskip
\centerline{The progression from Einstein-K\"ahler to Einstein-Sasaki
to Ricci-flat K\"ahler cone.}
                 
\bigskip\bigskip

\subsection{Lichnerowicz bound for Einstein-Sasaki spaces}

    In any Einstein space $M$ that admits Killing spinors, we can
prove that the bound (\ref{instability}) that governs the stability of
AdS$\times M$ solutions, and also the stability of
Schwarzschild-Tangherlini black holes, is always satisfied. In other
words, we can prove that an Einstein space in $d$ dimensions with
cosmological constant $\Lambda$ has a Lichnerowicz spectrum such that
\be
\Delta_L \ge  \fft{\Lambda}{d-1}\,\Big( 4 - \frac{(5-d)^2}{4}\Big)\,. 
\label{instability2}
\ee

   To prove this we shall first, for convenience, make our conventional 
choice of normalisation $\Lambda=d-1$.  A Killing spinor therefore
satisfies $D_a\, \eta = \ft12 \im\, \Gamma_a\, \eta$.  Suppose that
$h_{ab}$ is a transverse traceless mode of the Lichnerowicz operator
on $M$:
\be
\Delta_L\, h_{ab} = \lambda\, h_{ab}\,,\qquad \nabla^a\, h_{ab}=0\,,\qquad
h^a{}_a=0\,.
\ee
We now define two vector-spinors:
\be
\phi_a\equiv h_{ab}\, \Gamma^b\,\eta\,,\qquad \chi_a\equiv
(\nabla_b h_{ac})\, \Gamma^{bc}\, \eta\,.
\ee
The assumed properties of $h_{ab}$ can easily be seen to imply that
\be
D^a\, \phi_a=0\,,\quad \Gamma^a\, \phi_a=0\,,\quad D^a\, \chi_a=0\,,\quad
\Gamma^a\, \chi_a=0\,.
\ee

   We now calculate the action of the Rarita-Schwinger operator on
the vector-spinors, finding after some algebra that
\be
\im\, \Gamma^b\, D_b\, \phi_a= \chi_a -\ft{\im}{2}\, (d-2)\, \phi_a\,,
\qquad
\im\, \Gamma^b\, D_b\, \chi_a= -(\lambda -d)\,\phi_a + \ft{\im}{2}\, 
(d-4)\, \chi_a\,.\label{lemmata}
\ee
Thus by taking an appropriate linear combination of the two vector-spinors,
we can form an eigenfunction $\psi_a=\phi_a + k\, \chi_a$ of the 
Rarita-Schwinger operator on transverse gamma-traceless spin $\ft32$
modes,
\be
\im\, \Gamma^b\, D_b\, \psi_a = \mu\, \psi_a\,.
\ee
It follows immediately from (\ref{lemmata}) that we shall have an 
eigenfunction if
\be
\mu= \ft12 (d-2) -\im\, k\, (\lambda-d)\,,\qquad 
k\, \mu=\im -\ft12 k\, (d-4)\,.
\ee
These equations determine the constant of proportionality to be
$k= \im\, /[\mu + \ft12(d-4)]$, and hence that the Rarita-Schwinger
eigenvalue $\mu$ satisfies
\be
4\mu^2-4\mu -d^2 +10d -8 = 4\lambda\,.
\ee
Reorganising this we obtain
\be
\lambda= \ft14 (2\mu-1)^2 + 4 -\ft14 (d-5)^2\,.
\ee
From the reality of the Rarita-Schwinger eigenvalue $\mu$, we therefore
deduce that
\be
\lambda\ge  4 -\ft14 (d-5)^2\,.
\ee
Restoring the cosmological constant, we therefore obtain the claimed 
inequality (\ref{instability2}), which must hold for any Einstein
space of positive Ricci tensor that admits Killing spinors.  In particular,
this encompasses the case of all Einstein-Sasaki manifolds, in all 
odd dimensions.

   It is worth remarking that the above proof is a generalisation of
an argument that was used in \cite{dnp} in the case of
seven-dimensional Einstein-Sasaki manifolds.  It was argued there that
such a manifold $M_7$ could be used in order to obtain a
supersymmetric solution AdS$_4\times M_7$ of eleven-dimensional
supergravity.  Now it is known that eigenfunctions of the Lichnerowicz
operator in the internal space give rise to scalar fields in the AdS
spacetime.  The supersymmetry of the background implies that these
scalars must be members of supermultiplets, including fermions.  Since
the Kaluza-Klein reduction must necessarily give rise to {\it real}
masses for the fermions, it follows that the masses of the bosons (making
due allowance for the need to define mass carefully in AdS) must be
real also.  This translates into the statement \cite{dnp} that the
$(mass)^2$ of the scalars must respect the Breitenlohner-Freedman
\cite{bf} bound for stability, and hence it follows that the spectrum
of the Lichnerowicz operator must be bounded from below by the stability
limit, as given in (\ref{instability}), for the case $d=7$.  The same
argument was used recently for AdS$_5\times M_5$ compactifications
in \cite{dfghm}.  Of course our general proof above can be seen to 
be essentially an extension of the supersymmetry argument of
\cite{dnp}, since in fact the crucial ingredient was not really
supersymmetry {\it per se}, but rather, the fact that the mass spectrum
of scalar fields can be related to the mass spectrum of spin $\ft12$ 
fields.  Since the Lichnerowicz Laplacian is the mass operator for
scalar fields, and the Rarita-Schwinger operator is the mass operator for
spin $\ft12$ fields, our demonstration above that the eigenfunctions of
the two operators are related when there are Killing spinors can be
seen to reduce to the supersymmetry argument in those special dimensions
where supersymmetric AdS vacua can be found.  Our argument above is much
more general, however, since it dispenses with the excess baggage of
supersymmetry, and the need to interpret mass in AdS backgrounds.

   In \cite{pagepope,pagepope2} the Lichnerowicz bounds were
investigated for 7-dimensional Einstein metrics on the spaces $M(m,n)$
and $Q(k,\ell,m)$ which are $U(1)$ bundles over $\CP^2\times S^2$ and
$S^2\times S^2\times S^2$ respectively, with the integers specifying
the winding numbers of the $U(1)$ fibres over the base components.  It
is was found that for the Einstein-Sasaki examples, namely $M(3,2)$
and $Q(1,1,1)$, the bound $\Delta_L \ge \ft12\Lambda$ in
(\ref{instability2}) is strictly exceeded.  This has the interesting
consequence that for a range of ratios $m:n$ or $k:\ell:m$ around the
Einstein-Sasaki values, the stability bound is still satisfied despite
the absence of supersymmetry \cite{pagepope,pagepope2}.  By contrast,
it was shown recently in an analogous 5-dimensional calculation for
the $T^{p,q}$ spaces with Einstein metrics that the bound $\Delta_L
\ge \Lambda$ in (\ref{instability2}) is exactly saturated by the
Einstein-Sasaki case $T^{1,1}$, and that all the non-supersymmetric
$p\ne q$ spaces have a Lichnerowicz mode lying strictly below the 
bound \cite{dfghm}.

   It is worth remarking that, in view of the equivalence of the
criteria for black hole stability and AdS stability described in
sections \ref{blacksec} and \ref{adssec}, we have the immediate
consequence that Einstein-Sasaki manifolds will always give stable
Schwarzschild-Tangherlini black holes.

   A further consequence is that any Einstein metric whose
Lichnerowicz spectrum does not respect the lower bound
(\ref{instability2}) cannot admit Killing spinors, and so it cannot give
rise to supersymmetric backgrounds in any supergravity compactification.
Examples include not only the case of product metrics, for which
it has long been known that there exists a Lichnerowicz zero mode
\cite{dnp}, but also cases such as the Bohm metrics whose negative
Lichnerowicz modes we have demonstrated in this paper.

\section{Lorentzian Bohm Metrics, Real Tunneling Geometries,
and Counterexamples to the Cosmic Baldness Conjecture}

    In this section we discuss metrics obtained by analytic
continuation of the Bohm metrics. These metrics, which provide
generalisations of de Sitter spacetime as locally static solutions
with cosmological horizons, have a number of applications. In
particular, they provide counterexamples to a certain form of the
Cosmic Baldness conjecture. Furthermore, the Riemannian Bohm metrics have
a totally geodesic hypersurface. This allows them to be viewed as real
tunnelling geometries for the creation of the Lorentzian Bohm metrics
``from nothing.'' We first review the geometry by discussing the case of
the round $S^5 = \rm{Bohm}(2,2)_0$.

\subsection{Round $S^5$ and $dS_5$}\label{ds5sec}

   The round metric on $S^5$ may be written as
\be\label{eq:s5metric}
ds^2 = d\rho ^2 + \sin^2\rho \,\bigl ( d \theta ^2 + \sin^2 \theta d \phi
^2 \bigr )+ \cos^2\rho \,\bigl ( d {\theta ^ \prime}  ^2 + \sin ^2
{\theta ^ \prime } ^2 d {\phi ^\prime }^2  \bigr ) \,.
\ee
This provides an isometric embedding into ${\Bbb  E} ^6$, with Cartesian
coordinates denoted by the variables 
$( X_1, X_2, X_3, {X^ \prime}_1 , {X^\prime }_2, {X
^ \prime}_3 )$, via
\be\label{eq:embed1}
(X_1, X_2, X_3)= \sin\rho \, ( \sin \theta \cos \phi, \sin \theta \sin
\phi, \cos \theta)\,,
\ee
and
\be\label{eq:embed2}
({X ^\prime}_1, {X ^ \prime}_2, {X
^ \prime}_3)= \cos\rho\,  ( \sin \theta ^\prime \cos \phi ^\prime , \sin
\theta ^\prime  \sin \phi ^\prime  , \cos \theta ^\prime )\,.
\ee
Thus, as required, one has
\be
X_1^2 + X_2^2 +  X_3^2 + {X^\prime}_1^2
+ {X^\prime}_2^2 + {X^\prime}_3^2 = 1\,.
\ee
Note also, for later use, that
\be\label{eq:spheres}
X_1^2 + X_2^2 + X_3^2 = \sin^2\rho \,,\qquad
{X^\prime}_1^2 + {X^\prime }_2^2 + {X^\prime}_3^2 = \cos^2\rho \,.
\ee
The range of $\rho$ is seen to be $\rho  \in  [0, {\pi \over 2} ]$.

   To get the locally static Lorentzian de Sitter solution $dS_5$, one
can set the angle $\phi= \im\, t$ with $t$ real.  This means that $
X_2= \im\, T$ with $T$ real, and the embedding is into (but not onto)
the quadric
\be
X_1^2 - T^2 + X_3^2 + {X^\prime}_1^2 +
{X^\prime}_2^2 + {X^\prime}_3^2 = 1\,.
\ee
Equations (\ref{eq:spheres}) now become
\be
X_1^2 - T^2 + X_3^2 = \sin^2\rho \,,\qquad 
{X^\prime}_1^2 + {X^\prime}_2^2 + {X^\prime}_3^2 =  \cos^2\rho \,.
\ee
It is clear that there can be points on $dS_5$ for which $
X_1^2 - T^2 +  X_3^2$ is negative, and therefore for
which $ {X^\prime}_1^2 + {X^\prime}_2^2 +
{X^\prime}_3^2 $ exceeds unity. It follows that we need
to use a different parameterisation. We set $\sin^2\rho = 1-b^2$ and
get the metric
\be\label{eq:ds5metric}
ds^2 = \frac{db^2}{1-b^2} + (1-b^2 )\bigl ( d\theta^2 -\sin^2
\theta dt^2 \bigr )+ b^2 \, \bigl ( d {\theta^\prime}^2 +
\sin^2 \theta^\prime d{\phi^\prime }^2  \bigr )\,.
\ee
In this metric $b \in [0,\infty)$. However, $b=1$ is a coordinate
singularity, and for $b>1$ the orbits of $\partial_t$ are
spacelike. We use (\ref{eq:ds5metric}) in the region $0\leq b < 1$,
where $\partial_t$ is timelike.  There are Killing horizons of
$\partial_t$ at $\theta=0$ and $\theta=\pi$. The metric on the horizon
is
\be
ds^2 = \frac{db^2}{1-b^2} + b^2 \, \bigl ( d {\theta^\prime}^2 +
\sin^2 \theta^\prime d{\phi^\prime }^2  \bigr ) \,,
\ee
which is the standard metric on $S^3$. It is important to note that 
in order to
get all of the $S^3$ we need both of the copies of this metric
with $0 \leq b < 1$ that from each of the two values $\theta=0$ {\it and}
$\theta=\pi$, which each cover half of the horizon. This is best seen
from the embedding of (\ref{eq:embed1}) and (\ref{eq:embed2}). The
cosmological event horizon is not unique, since one may act with the
$SO(5,1)$ isometry group. This fact is connected with the observer
dependence of the associated Hawking thermal radiation
\cite{GibbonsHawking1}. The coordinate singularity at $b=1$ arises
because the 2-dimensional orbit of $SO(2,1)$ on the $dS_2$ factor
changes from being timelike to spacelike as it crosses the surface
$b=1$.

   It is interesting to observe that with the metric on $S^5$ written as
in (\ref{eq:s5metric}), the map $\phi \rightarrow - \phi$ is an
isometry, which fixes pointwise a separating totally-geodesic hypersurface
$\Sigma$, given by $\phi=0$ {\it and} $\phi=\pi$ (because $-\pi \sim
\pi$ here).  In terms of the embedding described earlier, $\Sigma$ is
the hypersurface $X_2=0$.  The ${\mathbb{Z}}_2$ isometry implies the
vanishing of the second fundamental form on $\Sigma$, \ie $K \equiv
\frac{1}{2}{\mathcal{L}}_n g = 0$, and hence the totally-geodesic 
property.  Here $n$ is the normal to $\Sigma$.  The metric on $\Sigma$ is
\be\label{eq:s4metric}
ds^2 = \frac{db^2}{1-b^2} + (1-b^2 ) \, d\theta^2 +
b^2 \, \bigl ( d {\theta^\prime}^2 +
\sin^2 \theta^\prime d{\phi^\prime }^2  \bigr )\,,
\ee
which is in fact just the round metric on $S^4$. We get all of $S^4$
because we have two copies of this metric, each with $\theta\in[0,
\pi]$, corresponding to $\phi=0$ and $\phi=\pi$.  Again, this is seen
most immediately in terms of the embedding (\ref{eq:embed1})and
(\ref{eq:embed2}), where $X_2=0$ manifestly defines an $S^4$.  Thus we
have a real tunnelling geometry in the sense of
\cite{GibbonsHartle}. That is, we have a compact gravitational
instanton with totally-geodesic boundary, such as one might use to
approximate a proposed wave function for the universe. The Riemannian
metric may be grafted onto the Lorentzian $dS_5$ metric
(\ref{eq:ds5metric}) at $t=0$, where it is clear that
(\ref{eq:ds5metric}) also has the same totally-geodesic hypersurface
with metric(\ref{eq:s4metric}) defined as the fixed point set of $t\to
-t$.

\subsection{Lorentzian Bohm metrics}

   The setup of the previous subsection \ref{ds5sec} generalises
straightforwardly to Bohm metrics. An isometric embedding is no longer
possible, because spheres are the only positive curvature Einstein
metrics that may be embedded isometrically into Euclidean space of one
dimension higher.  However, the topological statements go through. The
analytic continuation of the five dimensional Bohm metrics gives
\be\label{eq:lorentzbohm}
ds^2 = d\rho ^2 + a^2(\rho) \,\bigl ( d \theta ^2 - \sin^2 \theta dt
^2 \bigr )+ b^2(\rho) \,\bigl ( d {\theta ^ \prime}  ^2 + \sin ^2
{\theta ^ \prime } \,d {\phi ^\prime }^2  \bigr ) \,.
\ee
Note that we could have analytically continued the second sphere
instead.  In some cases this gives two inequivalent Lorentzian
metrics; we shall discuss making this interchange $a \leftrightarrow
b$ below.  Again, we have Killing horizons at $\theta=0$ and $\theta =
\pi$.  The range of $\rho$ depends on the specifics of the Bohm metric.

    Consider first the cases where the corresponding Riemannian Bohm
metric has topology $S^5$. The topology of the horizon is $S^3$, as it
was for the round $S^5$ of the previous subsection \ref{ds5sec}.  This
follows from the fact that the metric functions $a(\rho)$ and $b(\rho)$
behave near the two endpoints $\rho=0$ and $\rho=\rho_f$ in the same
way as $\sin\rho$ and $\cos\rho$ respectively behave near the
endpoints $\rho=0$ and $\rho=\frac{\pi}{2}$ of the round $S^5$ metric.
As we shall see shortly, the area $A$ of this cosmological event
horizon is always less than in the round case.  Interchanging $a$ and
$b$ in the $S^5$ Bohm examples gives the same Lorentzian metric.

    When the Riemannian Bohm metric has topology $S^3\times S^2$, an
exchange of functions $a \leftrightarrow b$ in (\ref{eq:lorentzbohm})
will change the topology of the Lorentzian manifold, and in particular
the topology of the event horizon. This is because in these cases
$a(\rho)$ goes to zero at both endpoints, whilst $b(\rho)$ never goes
to zero. For topological purposes, we may think of $a$ as behaving
like $\sin\rho$ with endpoints $\rho=0,\rho=\pi$ and $b$ behaving like
a constant function (just as in the ``trivial'' Bohm metric
Bohm$(2,2)_1$, which is simply the product Einstein metric on $S^3\times
S^2$). Thus we have two possibilities.  The metric
(\ref{eq:lorentzbohm}) has a horizon with topology $S^1\times S^2$. If
we exchange $a$ and $b$, the horizon will have topology $S^3$.  These
topologies are seen in the same way as in the previous subsection
\ref{ds5sec}, and as always we should take care to include the two
values $\theta=0$ and $\theta=\pi$.  The area of the horizons are
\be
A_1 = 8 \pi \int a^2 dt \,
\ee
and
\be
A_2 = 8 \pi \int b^2 dt \,,
\ee
respectively. Note that there is an extra factor of 2 because there
are contributions from both $\theta=0$ and $\theta=\pi$.  The
arguments of the following sections suggest that these two quantities
should be equal, and less than the horizon area for the de Sitter
spacetime $dS_5$.  These are nontrivial conditions on the functions
$a$ and $b$.

    The non-uniqueness of these cosmological horizons is reduced
compared to the de Sitter case, because the relevant isometry group is
now only $SO(2,1)$.

    The ${\mathbb{Z}}_2$ isometry of the previous subsection
\ref{ds5sec} is also present in the Bohm metrics, and therefore we
recover a totally-geodesic submanifold. Thus one might consider using
Bohm metrics in tunnelling calculations for the creation of a
Lorentzian Bohm universe.  In that application the number of negative
modes of $\Delta _2$ should be an odd number, so as to get an
imaginary part for the free energy when one evaluates the functional
integral.  Usually one expects just one negative mode, and the
contribution of instantons with more than one is often ignored.

    Another interesting question is what is the volume of $\Sigma$,
the totally-geodesic boundary.  For tunnelling geometries
constructed from hyperbolic tunnelling manifolds \cite{GibbonsB}
\footnote{Note that $\Delta_2$ has no negative modes in this case, and
so one might worry about the tunnelling interpretation.}, 
${\rm Vol} (\Sigma)$ is a measure of the complexity of
$\Sigma$ and it is possible to bound the volume of the tunnelling
geometry in terms of the volume of the boundary such that larger
complexity, \ie larger ${\rm Vol} (\Sigma)$, means larger volume
\cite{GibbonsC}. In the case of positive scalar curvature, and with
boundaries $\Sigma$ of simple topology ( $S^4$ or $S^2 \times S^2$ in
our case), the notion of complexity is not relevant.  However, it is still
interesting to know how the volume of the manifold is related to the
volume of the totally-geodesic boundary. Specifically, the volume of
the five-dimensional manifold is
\be
V = 16 \pi ^2 \int a^2\, b^2 \, dt\,,
\ee
while
\bea
{\rm Vol}(\Sigma_2) = 8 \pi ^2 \int a^2 \, b \, dt , \nonumber \\
{\rm Vol}(\Sigma_1) = 8 \pi ^2 \int a \, b^2 \, dt \,.
\eea
The two different values for $\Sigma$ correspond to interchanging $a$
and $b$ in the metric. In the $S^5$ cases these will be the same, but
for the $S^3\times S^2$ cases they will be different. Some examples
are illustrated in the following table. The table also collects
information about the corresponding values of $b_0$ and $t_{fin}$.

\begin{table}[ht]
  \begin{tabular}{|c|c|c|c|c|c|c|} \hline
Bohm & Topology & $b_0$ & $t_{fin}$ & V = &
${\rm Vol}(\Sigma_2) =$ & ${\rm Vol}(\Sigma_1) = $ \\
 & & & & $16 \pi^2 \int a^2 b^2 dt$ & $8 \pi^2 \int a^2 b dt$ &
$8 \pi^2 \int a b^2 dt$ \\ \hline \hline
$(2,2)_0$ & $S^5$            & 1        & 1.57079 & 31.006 & 26.320 
& 26.320 \\
\hline
$(2,2)_2$ & $S^5$            & 0.253554 & 2.68470 & 20.814 & 20.302 
& 20.302 \\
\hline
$(2,2)_4$ & $S^5$            & 0.053054 & 3.04979 & 20.672 & 20.2605 
& 20.2605 \\ \hline \hline
$(2,2)_1$ & $S^3 \times S^2$ & 0.5      & 2.22143 & 21.924 & 21.924 
& 19.739 \\
\hline
$(2,2)_3$ & $S^3 \times S^2$ & 0.117794 & 2.93537 & 20.684 & 20.189 
& 20.335 \\
\hline
$(2,2)_5$ & $S^3 \times S^2$ & 0.023571 & 3.10092 & 20.6709 & 20.267 
& 20.254 \\ \hline
$(2,2)_{\infty}$ & bi-cone on $S^2 \times S^2$ & 0 & 3.14159 
& 20.6708 & 20.2603 & 20.2603\\ \hline
  \end{tabular}
\end{table}

    In this table the results for the double cone are found
analytically, and the Bohm$(2,2)_0$ metric (the round 5-sphere) may also be
calculated analytically as a check on the numerics. The volumes decrease
from the round sphere to the double cone, as expected from Bishop's
theorem. The volumes of the $\Sigma$'s decrease with the volume in the
$S^5$ cases, but not in the $S^3\times S^2$ cases.

    Similar calculations may of course be done with higher-dimensional
Bohm metrics, as illustrated in the following table. The
generalisation of $\Sigma$ to higher dimensions is slightly more
involved, and will not be discussed here.

\begin{table}[ht]
\centerline{
  \begin{tabular}{|c|c|c|c|c|} \hline
Bohm & Topology & $b_0$ & $t_{fin}$ & V =  \\
 & & & & $4 \pi^4 \int a^3 b^3 dt$ \\ \hline \hline
$(3,3)_0$ & $S^7$            & 1 & 1.5707 & 32.470  \\ \hline
$(3,3)_2$ & $S^7$            & 0.305521       & 2.6933 
& 24.499 \\ \hline \hline$(3,3)_1$ & $S^4 \times S^3$ & 0.577351 
& 2.2207 & 24.995  \\ \hline
$(3,3)_3$ & $S^4 \times S^3$ & 0.14291   & 2.9322 & 24.482 \\ \hline
$(3,3)_{\infty}$ & bi-cone on $S^3 \times S^3$ & 0 & 3.1416 
& 24.4816 \\ \hline
  \end{tabular}}
\end{table}

\subsection{Cosmological event horizons}

    In the Riemannian metrics, the circle action on $S^5$ generated by
$\partial _\phi$ rotates the $X_1-X_2$ plane. The action has an
$S^3$'s worth of fixed points for which $X_1=0$ and $X_2=0$, corresponding to
$\theta=0$ {\it and} $\theta=\pi$. Because the reversal of $\phi$ is also
an isometry, we have in fact an $O(2)$ action, which allows the analytic
continuation to a locally static, \ie  time-reversal invariant,
metric with a hypersurface-orthogonal locally timelike Killing vector
field with a Killing horizon.  These are the cosmological horizons of
the previous subsections.  In such cases the Lorentzian metric may be written
locally as
\be
ds^2 = - U^2 \, dt ^2 + g_{ij} \, dx ^i\,  dx ^j\,,
\ee
where $g_{ij}$ is the metric on the orbit space Q of $\partial_t$,
and the gravitational field equations imply in particular that
\be
\nabla ^2 U= - \Lambda U\,.
\ee

    The quantity $U$ vanishes on the horizon, and its normal
derivative $ \partial U \over \partial n$ on the horizon $U=0$ is a
constant, which is called the surface gravity $\kappa$.  The period in
imaginary time, \ie the real period of $\tau \equiv \im \,t$, is $2
\pi/\kappa$. If $V$ is the volume of the corresponding
Riemannian manifold, and $A$ is the area of the event horizon in the
Lorentzian manifold, one has 
\be V= { 2 \pi \over \kappa}\,  \int_Q U\, 
\sqrt g \, d^{n-1} x \,.  
\ee 
The boundary $\partial Q$ is the event horizon, where the orbits
degenerate. For example, when the Bohm metric has topology $S^5$ the
boundary $Q$ is a 4-ball, $B^4$, with boundary the even horizon
$S^3$. The metric on $Q$ is in fact given by
(\ref{eq:s4metric}). However, the crucial difference is that we take
only $\phi=0$ to intersect all the orbits of $\partial_{\phi}$ once,
whilst for $\Sigma$ we needed to take both $\phi=0$ and
$\phi=\pi$. The $S^4$ that we had before corresponded to gluing two
copies of $B^4$s across their boundary $S^3$.

   Integration of $\nabla ^2 U$ gives
\be
\Lambda \int_Q U \, \sqrt g \, d^{n-1} x = \kappa\,  A\,,
\ee
whence
\be\label{eq:VandA}
V \, \Lambda =  2 \pi \, A\,.
\ee
This argument shows  that when there are two possible inequivalent
analytic continuations, such as for the Bohm metrics on $S^3\times
S^2$, the horizon areas should be the same, $A_1 = A_2$. This is
illustrated in the following table, which also illustrates the
relationship (\ref{eq:VandA}), showing that it works for the various
topologies.  The values of $V$ are repeated from the previous table.

\begin{table}[ht]
\centerline{
  \begin{tabular}{|c|c|c|c|c|c|} \hline
Topology & $b_0$ & $A_1$ &
$A_2 = A_1$ ? & $V$ & $4V = 2\pi A$ ? \\ \hline \hline
$S^5$            & 1        & 19.74 & Yes & 31.006 
& Yes $(=124.02)$ \\ \hline
$S^5$            & 0.253554 & 13.25 & Yes & 20.814 
& Yes $(=83.25)$ \\ \hline
$S^5$            & 0.053054 & 13.160 & Yes & 20.672 
& Yes $(=82.688)$ \\ \hline
\hline
$S^2 \times S^3$ & 0.5      & 13.96 & Yes & 21.925 
& Yes $(=87.69)$ \\ \hline
$S^2 \times S^3$ & 0.117794 & 13.168 & Yes & 20.684 
& Yes $(=82.74)$ \\ \hline
$S^2 \times S^3$ & 0.023571 & 13.15954 & Yes & 20.6709 
& Yes $(=82.684)$ \\ \hline
bi-cone on $S^2 \times S^2$ & 0 & 13.15948 & Yes & 20.6708 
& Yes $(=82.683)$\\ \hline
  \end{tabular}}
\end{table}

    Note that $\kappa$ cancels in (\ref{eq:VandA}), as it must since
it depends on the normalisation of the length of the Killing field,
which is arbitrary. This relation between area and volume is quite
{\sl universal} and holds for any Einstein metric admitting an $O(2)$
action. It allows us to relate the on-shell action to the area of the
horizon, and hence to show that formally at least, the entropy $S$ is
given by
\be
S= \ft14  A\,,
\ee
just as in four dimensions. This general argument was first given
in four dimensions in \cite{GibbonsHawking2}.  Now, a theorem of Bishop
\cite{Bishop} tells us that for fixed $\Lambda$, the volume $V$ never
exceeds its value for the round sphere, with equality only in the case
of roundness. It follows that this area or entropy is always less than 
the area or entropy of the corresponding horizon in $dS_5$.

    Since dynamically one expects the area to increase, and
thermodynamically one expects the entropy to increase, there seem to
be some physical grounds for believing that the static Lorentzian Bohm
metrics are dynamically unstable. Indeed, one might conjecture that if
they are perturbed slightly at some initial time, then they will
evolve to an asymptotically de Sitter state, and that this evolution
will be such that the area of the cosmological horizon increases
monotonically from a value near, but smaller than, its value for the
initial Bohm metric, to its value for the de Sitter spacetime. It
should not be impossible to investigate this conjecture numerically.

\subsection{Consequences for Cosmic Baldness}

     It has been conjectured for some time \cite{BoucherGibbons} (in
four dimensions) that there should exist only one regular static
solution of the Einstein equations with a cosmological constant that
has only a single cosmological horizon (so that $\partial Q =
S^{n-2}$). In fact one usually thinks of $Q$ as being topologically an
$(n-1)$-ball, as described above. This is a much stronger statement
than that locally, within the event horizon of every observer or most
observers, the metric will settle down to the static de Sitter
form. In fact for generic initial data one cannot hope that the metric
will settle down globally to the de Sitter state, as was originally
made clear in \cite{BoucherGibbons} (see also \cite{GibbonsRuback} for
a detailed discussion using the exact Lorentzian Taub-NUT metrics).

    It is now clear that the Bohm metrics provide infinitely many
counterexamples to the Cosmic Baldness conjecture in dimensions
$5\le n\le 9$. The situation in four dimensions remains unclear.  It
is still possible to believe an even stronger conjecture, the truth of
which would imply the Cosmic Baldness conjecture, namely that there is
only one Einstein metric on $S^4$.  At present all that is known is
that if there is another Einstein metric on $S^4$, then its volume
must be less than that of the round metric by a factor of 3
\cite{Gursky}, and that the magnitude of the Weyl tensor must exceed a
certain threshold \cite{Singer}. This is interesting in the light of
the fact that it is the magnitude of the Weyl tensor which appears to
play a role in controlling the spectrum of the Lichnerowicz
operator. It may perhaps suggest that any counterexample will have a
negative mode of $\Delta_2$.

    Curiously, there are some proofs of a form of the Cosmic No-Hair
conjecture in the literature \cite{Boucher, Friedrichs}, but these
proofs require a smooth structure at future spacelike infinity
$I^+$. It seems likely that in our examples, the future timelike
infinity will not be of the sort envisaged in those proofs.  It would
be interesting to investigate this point further, but this would seem
to require analytic formulae for $a(\rho)$ and $b(\rho)$.  As
mentioned in the previous sub-section it seems likely that these
static metrics will be dynamically unstable, and they may well evolve
into an asymptotically de Sitter-like state. If this is true then the
main physical spirit of the No-Hair conjecture will hold, even though
the letter of the Baldness conjecture is broken.

\section{Negative modes and non-uniqueness of the
Dirichlet problem}

    The existence of negative Lichnerowicz modes has been a major
theme in this work.  This section contains speculative comments on a
generic connection between negative Lichnerowicz modes and the
non-uniqueness of solutions to the Dirichlet problem.  In particular,
we argue that there will be infinitely many negative Lichnerowicz
modes on the {\it non-compact} Ricci-flat Bohm metrics.  The existence
of $L^2$ negative modes for the Lichnerowicz operator for non-compact
Ricci flat manifolds such as the Riemannian Schwarzschild solution
first came to light when considering the negative specific heat of
black holes. Since then a considerable literature has grown up,
analysing that and related cases. 

    In the non-compact Ricci flat case, it seems that the general
picture is as follows. One has a class of metrics on a manifold $M_n$
depending on some parameters, in the simplest case just one overall
scaling parameter $\mu$, such as the mass in the Schwarzschild
case. One asks whether this metric can fill in a given boundary
$\Sigma _{n-1}$ that has a given metric $h_{ij}$. That is, one tries to solve
the Dirichlet problem for the Einstein equations.  In the
four-dimensional Riemannian Schwarzschild case for example, the boundary
is taken to be $S^2 \times S^1$ with the product metric. This is specified
by the radius $R$ of the 2-sphere and the period $\beta$ of the
circle. Physically, we are putting a black hole in a spherical box of
radius $R$, and fixing the temperature on the boundary of the box to
equal $T= \beta ^{-1}$.  If the metric and the boundary data are to
agree then we must have
\be
8 \pi \, \mu  \,  \sqrt{1 - \frac{2\mu}{R} } =\beta\,.
\ee
The number of solutions of this equation for $\mu$ depends on the
ratio $\beta/R$ that specifies, up to a scale, the boundary metric
$h_{ij}$.  One finds that if the ratio is small there are two solutions
for $\mu$.  The Einstein action $I$ of the two solutions differs.  The
action for the smaller value of $\mu$ is the smaller.  We shall refer to these
two solutions as branches.  If the ratio is large there are no
solutions for $\mu$.  At the critical value the two solutions for
$\mu$ coincide, and give $\mu = \ft13 R$.

    Now consider the operator $\Delta_2$ (which equals $\Delta_L$ in
this Ricci-flat situation), subject to Dirichlet boundary conditions,
which gives the Hessian of the action $I$.  For a large box (in
relation to the scale set by the temperature $T= \beta ^{-1}$),
$\Delta_2$ has a single negative mode for the branch with the smaller
value of $\mu$, and it a positive but no negative mode for the branch
with the larger value. As one reaches the critical value, the two
branches coincide and so do the two eigenvalues.  At the critical
point there is an eigenmode of $\Delta_L$ with zero eigenvalue. In
other words, at the critical point there is a marginally stable
mode. Because the specific heat is given essentially by the Hessian of
the action (\ie the free energy) considered as a function of the
boundary data, it changes sign at this value. 

   Next, consider what happens as the radius of the box increases to
infinity, so that the ratio $\beta/ R$ goes to zero while keeping $\beta$
a constant, and take the branch on which $\mu$ is smaller. In
the limit, one finds that $\mu \rightarrow \beta/(8 \pi)$. On
the other branch, one finds that $\mu \rightarrow \infty$. Thus, in the
limit $R \rightarrow \infty$ one gets a non-compact Ricci-flat
Einstein manifold, namely the standard Schwarzschild solution, and by
following the mode which first appears as a zero-mode at the critical
value, one gets an $L^2$ negative mode for $\Delta_L$ on that manifold. This
process is illustrated at fixed temperature in the following figure.

\begin{figure}[htbp]
\centerline{\epsfxsize=3.4truein
\epsffile{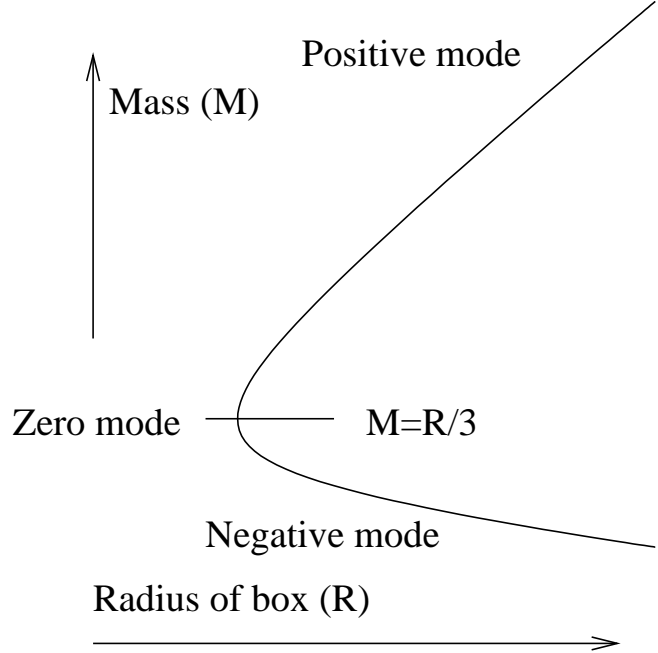}
}
\caption{Relationship between Lichnerowicz modes and masses for
Schwarzschild in a finite cavity.
}
\end{figure}

    The picture described above has been vindicated by detailed
numerical calculations in this and related cases 
\cite{Prestidge,GregoryRoss,GubserMitra1,GubserMitra2,Reall}.
For example, Hawking and Page \cite{HawkingPage} studied black holes
in anti-de Sitter spacetime. The classical solution is the Kottler or
Schwarzschild anti-de Sitter solution, an Einstein metric with
negative scalar curvature $4 \Lambda$. The role of the radius $R$ is
now played by the cosmological constant $\Lambda$, and the manifolds
considered are always non-compact, but the general picture is similar.

   The arguments given above are heuristic rather than being
completely rigorous, but they suggest the following
generalisation. One considers a one-parameter family of Dirichlet
problems for the Einstein equations. As the parameter varies one finds
a discrete non-uniqueness, with more and more branches appearing,
generically in pairs, and as each new branch appears a zero mode of
$\Delta_2$ occurs, which then splits into a pair of modes, one with
positive eigenvalue and one with negative eigenvalue. In the limit
that one gets a non-compact manifold, one should have found, on the
correct branch, as many $L^2$ negative modes as the number of critical
values one has passed.

    An obvious example on which to try this argument is the noncompact
metric of Bohm on ${\Bbb R}^3 \times S^2$ \cite{Bohm2}, recently
considered by Kol \cite{Kol} and discussed above. In fact it exhibits
a feature not seen previously, which is that even within the
restricted framework of cohomogeneity one metrics, the Dirichlet
problem may have infinitely many solutions.

    These metrics are determined by a single (scale) parameter, which
may be taken to be the radius $b_0$ of the 2-sphere bolt.  Now the
geometry of a boundary at some fixed value of the radius $R$ is given
by the ratio $a(R)/b(R)$. Thus the possible filling solutions are
determined by the intersections of the curve of $a/b$ with a straight
line through the origin in the $(a,b)$ plane, as shown in figure
11. Clearly, if we let $a/b$ tend to one there are more and more
intersections, which appear at critical points in pairs when the
straight line touches the slightly wiggly but almost straight $a/b$
curve. It seems reasonable to suppose that an additional zero-mode of
$\Delta_L$ appears at this point, and then as the slope of the
straight line gets closer to unity, a pair of eigenvalues, one
positive and one negative, branches off. If this intuition is correct,
and if the branches are connected at $a/b=1$ say, then one expects
that the noncompact Bohm metrics on ${\Bbb R}^3 \times S^3$ should
have infinitely many $L^2$ Lichnerowicz negative modes. This process
is illustrated in the following figure.

\begin{figure}[htbp]
\centerline{\epsfxsize=3.4truein
\epsffile{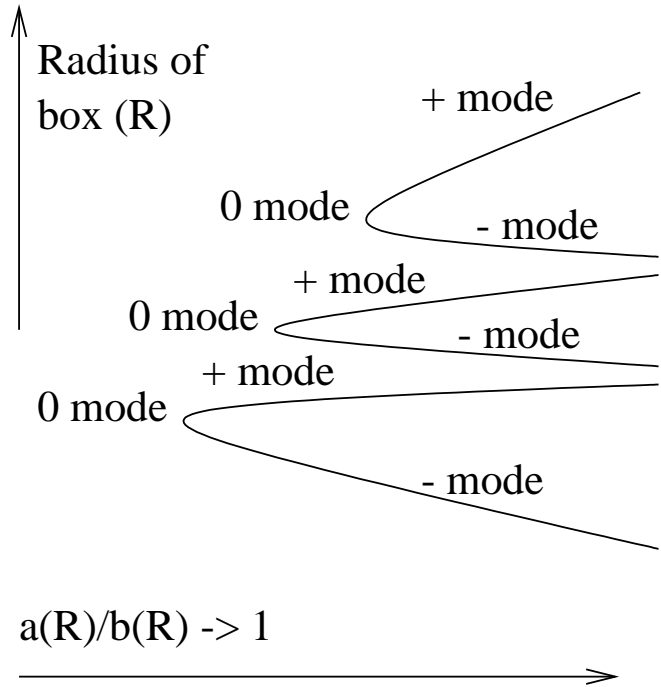}
}
\caption{The noncompact Bohm metric in a box with a fixed value of
$a/b$ at the boundary. Allowed values of $R$ are shown along with the
corresponding positive, negative and zero modes from branching. It is
expected that the branches will join at $a(R)/b(R)=1$.
}
\end{figure}

    Some evidence for this viewpoint comes from examining $SO(p)
\times SO(q)$-invariant transverse tracefree perturbations of the
singular Ricci-flat cone on $S^p \times S^q$ obtained in the limit
$b_0 \rightarrow 0$. That is to say, the cone is
\be
ds^2 = dr^2 + \frac{r^2}{p+q-1}\left((p-1) 
d\Omega^2_p + (q-1) d\wtd\Omega^2_q \right) ,
\ee
and the perturbation is
\be
h_{\alpha\beta} = r^2 \phi(r)
\left(
\begin{array}{cc}
\frac{1}{p} g_p & \\
 & - \frac{1}{q} \td g_{q} \\
\end{array}
\right)_{\alpha\beta} \,,
\ee
where $g_p$ and $\td g_q$ are the round metrics on $S^p$ and $S^q$,
respectively.  This is a zero mode on $S^p \times S^q$. If we
want a mode on the cone with Lichnerowicz eigenvalue $-\lambda$, the
equation for $\phi$ is \cite{gh}
\be
\frac{d^2\phi}{dr^2} + \frac{d}{r} \, \frac{d\phi}{dr} + 
\frac{2d-2}{r^2} \, \phi = \lambda \, \phi\,.
\ee
The solutions to this equation for $\lambda \neq 0$ are, writing $d=p+q$,
\be\label{eq:positive}
\phi(r) = r^{(1-d)/2} \left[A \, I_\mu 
(\lambda^{1/2} r)
+ B \, K_\mu (\lambda^{1/2} r) \right] \,,\qquad
\mu = \ft12 \sqrt{(d-1)(d-9)}\,,
\ee
where $A$ and $B$ are constants, and $I_{\mu}$ and $K_{\mu}$ are the
modified Bessel functions. When $\lambda = 0$ we get
\be\label{eq:zero}
\phi(r) = r^{(1-d)/2 \pm i \sqrt{(d-1)(9-d)} /2} .
\ee
This expression was written down in \cite{Kol} (his $d$ is shifted by
one).  It is also the behaviour of the Bessel function solutions in
(\ref{eq:positive}) as $r\to 0$. As $r\to\infty$, the Bessel function
solutions go as $r^{-d/2}\,  e^{\pm r}$. The $K_{\mu}$ Bessel function is
the better behaved.

    We are interested in the Bohm cases $4< d<9$ that
coincide with oscillatory behaviour in (\ref{eq:positive}) and
(\ref{eq:zero}).  For these dimensions, the zero-mode solutions are
not normalisable as $r\rightarrow \infty$ or as $r \rightarrow 0$,
although there are bounded as $r\to\infty$. The negative mode solution
with $K_{\mu}$ is normalisable at infinity.

     Thus the $K_{\mu}$ solutions decay at infinity, and as we move in
towards the origin, start oscillating at $r \sim
\lambda^{-1/2}$. In the singular cone limit, the modes are
not normalisable at the origin. However, suppose we are in a 
rounded-off cone. For $r \gg b_0$ the metric is essentially that of the
singular cone, and we may use our solutions (\ref{eq:positive}). If
further we have $\lambda^{-1/2}\gg b_0$, logarithmic oscillations
will set in, within this asymptotic regime. We should then expect to
be able to match this solution to a solution in the inner regions that
is well-behaved at the origin (cf. \cite{gh}), for a certain discrete
set of values for $\lambda$. This will give us a spectrum of negative
Lichnerowicz modes. It would seem that there will be an infinity of
such modes, accumulating at zero.

\section{Conclusions and discussion}

    The principal focus of this paper has been to study applications
of the countable infinities of inhomogeneous Einstein metrics on
certain spheres and products of spheres, which were discovered
recently by Bohm \cite{bohm}.  These occur for the topologies
$S^{p+q+1}$ and $S^{p+1}\times S^q$, for $5\le p+q\le 9$ and $p\ge 2$,
$q\ge 2$. They may be used in place of the usual round-sphere Einstein
metrics in a variety of constructions including black holes and
Freund-Rubin solutions, and after a Wick rotation to a Lorentzian
section, they may be interpreted as spacetime metrics in their own
right.

    The stability of generalised Schwarzschild-Tangherlini black
holes, where the $d$-dimensional constant-radius spatial sections
$M_d$ in the $(d+2)$-dimensional spacetime are taken to be positive
Ricci curvature Einstein spaces, was studied recently in \cite{gh}.
It was shown that a solution will be classically stable if the
spectrum of eigenvalues of the Lichnerowicz operator on transverse
traceless symmetric two-index tensors in $M_d$ is bounded below by a value
corresponding to $\Delta{\rm stab} \ge 0$ in (\ref{stabdef}). One of
our results in this paper has been to show that this stability
criterion is identical to one obtained in \cite{dnp,dfghm} for the
stability of Freund-Rubin solutions AdS$_n\times M_d$ of gravity
coupled to a $d$-form (or $n$-form) field strength.  Thus it becomes
of considerable interest to try to obtain bounds on the spectrum of
the Lichnerowicz operator on Einstein spaces $M_d$.

   The Bohm metrics are sequences of cohomogeneity one Einstein
metrics of the form (\ref{bohmans}), which more and more nearly
approach a ``double-cone'' form (\ref{bohmcone}) as one progresses 
along the sequence.  One therefore intuitively expects that the 
``ballooning'' instabilities associated with direct products of
the sphere metrics forming the principal orbits will give rise to
negative-eigenvalue modes of the Lichnerowicz operator.  Such modes
would {\it a fortiori} violate the stability criterion described
above, implying that Schwarzschild-Tangherlini black holes or
AdS$_n\times M_d$ solutions constructed using the Bohm metrics 
would be unstable.  In certain cases, including all the Bohm metrics
on $S^3\times S^2$ and $S^3\times S^3$, we constructed analytic
proofs that indeed show the existence of negative modes of the 
Lichnerowicz operator.  Numerical calculations for other examples,
namely Bohm metrics on $S^5$, have confirmed that these
too have negative-eigenvalue modes of the Lichnerowicz operator.
We believe that in fact all the Bohm metrics have negative 
Lichnerowicz modes.

   One can perform analytic coordinate continuations in the Bohm
metrics in order to obtain spacetimes with positive cosmological
constant that generalise de Sitter spacetime.  If one does this for
the Bohm metrics that are themselves topologically spheres, then the
resulting spacetimes have the same topology and global structure as de
Sitter spacetime itself.  These metrics provide infinitely many
counterexamples, in dimensions $5\le n \le 9$, to the Cosmic Baldness
conjecture, which asserted the uniqueness of regular static solutions
of the Einstein equations with a single cosmological horizon.
However, although the Bohmian analogues of de Sitter spacetime are
regular, we have argued that they are unstable and that they are 
likely to decay into a de Sitter-like state.  This would mean that
the No-Hair conjecture would remain inviolate.  

   In order to explore possible endpoints for the decay of spacetimes
constructed using Bohm metrics, we were also led to consider other
geometries for Einstein spaces $M_d$ that would satisfy the criteria
for stability.  In particular, we considered compact Einstein spaces of
positive Ricci curvature that admit Killing spinors.  We showed that
in all such spaces, in any dimension, one can derive a lower bound on
the spectrum of the Lichnerowicz operator which implies that the
stability criterion $\Delta_{\rm stab}\ge0$ is satisfied.  These
examples include all the Einstein-Sasaki spaces, which may be defined
as odd-dimensional Einstein spaces whose cones give Ricci-flat
K\"ahler spaces in one higher dimension.  It is straightforward to see
that the covariantly-constant spinors on the Ricci-flat K\"ahler cone
project down as Killing spinors on the Einstein-Sasaki base.  A
by-product of our results is that it demonstrates that the Bohm
metrics, for which $\Delta_L$ (and hence {\it a fortiori} $\Delta_{\rm
stab}$) can be negative, cannot admit Killing spinors.

   There also exist Ricci-flat Bohm metrics, with non-compact
topology.  The structure of these metrics at short distance looks very
similar to that near one of the two endpoints of the compact
metrics. However, lacking the cosmological term that causes the metric
functions to turn over and recollapse in the compact examples, the
non-compact spaces asymptotically approach the cone over the an
$S^p\times S^q$ direct-product base.  The non-compact spaces have
the topology $\R^{p+1}\times S^q$. The only parameter in the
non-compact metrics is the overall scale.  We constructed an analytic
proof that the non-compact Bohm metrics for $p=q=2$ and $p=2$, $q=3$ have
negative Lichnerowicz modes, and we presented a general argument that
indicates that all the non-compact Bohm metrics will have infinitely many
$L^2$ normalisable negative-eigenvalue Lichnerowicz modes.

   Another possible use of the Bohm metrics is to construct
four-dimensional gravitating monopoles and black holes by dimensional
reduction, as studied in \cite{hartnoll}. This is possible because
many of the Bohm metrics have $S^3$ factors.   Recalling that $S^3$
is isomorphic to $SU(2)$, one can quotient by a $U(1)$ action on 
$S^3$ to end up 
with $SU(2)/\left[ U(1)\times {\mathbb{Z}}_2 \right] \simeq
S^2$. Thus the resulting lower-dimensional space will have an $S^2$
factor, \ie it will look like a monopole or black hole, and it will come
with a $U(1)$ gauge field. Explicitly, one can write the metric on
$S^3$ using Euler angles as 
\be 
ds^2 = d\theta^2 + \sin^2\theta
d\psi^2 + (d\phi + \cos\theta d\psi)^2 \,, 
\ee 
where
$0\leq\theta\leq\pi$, $0\leq\phi < 2\pi$ and $0\leq\psi\leq 4\pi$. The
quotient by the $\partial_{\phi}$ isometry, and the ${\mathbb{Z}}_2$
quotient $\psi \sim \psi+2\pi$, leaves us with the standard metric on
$S^2$, namely $ d\theta^2 + \sin^2\theta d\psi^2$, and a charge-two Dirac
monopole $A = \cos\theta d\psi$. One can also quotient by the whole
$SU(2)$, and in this case because there is no fibration over the $S^3$
the $SU(2)$ gauge fields obtained will be trivial.

  Thus, for example, take the seven-dimensional noncompact
Bohm metric over two copies of $S^3$, supplement the metric by an eighth
timelike direction $-dt^2$, and dimensionally reduce on $SU(2)\times
U(1)$. One will obtain a gravitating $U(1)$ monopole with four scalar
fields. Another possibility would be to take the generalised black
hole in eight dimensions over a compact Bohm metric in six
dimensions with topology $S^3\times S^3$. Again quotient by
$SU(2)\times U(1)$, where the $SU(2)$ is acting on the round $S^3$ in
the Bohm metric. We will obtain a $U(1)$ magnetically charged black hole
in four dimensions. Because the $S^3$ that the $U(1)$ was acting on is
not round, the black hole will not be spherically symmetric.

\section*{Acknowledgements}

   We should like to thank Christoph Bohm for useful discussions.  
C.N.P. is grateful to the Isaac Newton Institute for
Mathematical Sciences and the Centre for Mathematical Studies in Cambridge,
and CERN, Geneva, for hospitality and financial support during the
course of this work.  C.N.P.~is supported in part by DOE
DE-FG03-95ER40917.  G.W.G.~acknowledges partial support from PPARC
through SPG\#613. S.A.H.~is funded by the Sims scholarship.

\vfill\eject

\centerline{\Large\bf APPENDICES}

\appendix
\section{Numerical Solutions for Bohm Metrics}

\subsection{Numerical techniques}

    The Einstein equations (\ref{einstein}) and (\ref{constraint}) 
cannot be solved explicitly when $p>1$ and $q>1$.  It was shown in
\cite{bohm} that countable infinities of smooth solutions satisfying 
the $S^{p+q+1}$ or $S^{p+1}\times S^q$ boundary conditions (\ref{bohmeven})
or (\ref{bohmodd}) exist.  It is quite a straightforward matter to 
obtain these solutions by numerical methods, since it turns out that
the second-order equations (\ref{einstein}) are quite stable.

   We have constructed numerical solutions by first obtaining Taylor-series
expansions for the metric functions $a$ and $b$ near to $t=0$, imposing
the $t=0$ boundary conditions given in (\ref{bohmeven}) (or, equivalently,
in (\ref{bohmodd})).  To the first couple of orders, these give
\bea
a &=& t -\fft{q\, (q-1) + b_0^2\, (p+q)(p-q+1)}{6b_0^2\, p\, (p+1)}\, t^3
  + O(t^5)\,,\nn\\
b &=& b_0 + \fft{q-1 - b_0^2\, (p+q)}{2b_0\, (p+1)}\, t^2 + O(t^4)\,.
\label{taylor}
\eea
Note that $a$ is an odd function of $t$, whilst $b$ is an even function.

   Using a Taylor expansion of the form (\ref{taylor}) (which we
actually evaluated up to order $t^9$), we then set initial data just
outside the singular point, for a very small positive value of $t$.
These data are then evolved forward in $t$ numerically, using the
second-order equations (\ref{einstein}).  The exercise then becomes a
``shooting problem,'' in which one seeks to adjust the one free
initial parameter $b_0$ so as to achieve a smooth termination of the
evolved data at a point $t=t_f$ where $a$ and $b$ satisfy one or other
of the $t=t_f$ boundary conditions given in (\ref{bohmeven}) or
(\ref{bohmodd}). 

    In cases where $p=q$, the numerical analysis is simpler, since the
regular solutions are all symmetric under reflection about the
midpoint $t=t_c=t_f/2$.  Thus one can avoid the need to handle the
integrations in the region near $t=t_f$ where one or other metric
function is tending to zero.  Instead, the shooting problem reduces to
finding a $b_0$ for which either $\dot a=\dot b=0$ at some point
$t=t_c$ (for the Bohm$(p,p)_{2m}=S^{2p+1}$ metrics), or else for which
$a=b$ and $\dot a = -\dot b$ (for the Bohm$(p,p)_{2m+1}=S^{p+1}\times
S^p$ metrics).

  It is known from the results in \cite{bohm} that
there is a countable infinity of values $b_0$ for which a regular 
termination at some $t_f$ occurs.  The largest $b_0$ yielding a regular
solution is $b_0=1$, leading to the standard unit $S^{p+q+1}$ metric
(\ref{bohm0}), which is called Bohm$(p,q)_0$.  The next value is $b_0=
\sqrt{(q-1)/(p+q)}$, giving the direct-product Einstein metric (\ref{bohm1})
on $S^{p+1,q}$ that we call Bohm$(p,q)_1$.  There is then a 
monotonically-decreasing sequence of $b_0$ values, giving the Bohm$(p,q)_n$
sequence of Einstein metrics, alternating between terminating $t_0$ boundary 
conditions given by (\ref{bohmeven}) and (\ref{bohmodd}).  The limit point 
of the sequence is $b_0=0$, giving the double-cone singular metric 
(\ref{bohmcone}).  

    Plots for the five-dimensional Einstein metrics Bohm$(2,2)_n$ on
$S^5$ and $S^3\times S^2$ are given below, for $0\le n\le 6$, with
$b_0=(1,\ft12,
0.253554255,0.117794,0.053054,0.023571,0.010503)$.  We also give a
plot for the case of Bohm$(2,3)_2$, which is topologically $S^6$, to
illustrate an example where there is no symmetry between $a$ an $b$.
This has $b_0\approx 0.297647$, and the endpoint is at $t_f\approx 
2.68296$ (there is no natural significance to the midpoint of the radial
coordinate range in the $p\ne q$ examples).

   A few isolated examples for other values of $p$ and $q$ are as follows.
The Bohm$(3,3)_2$ metric on $S^7$ has $b_0\approx 0.3055210896$, with the
midpoint occurring at $t=t_c\approx 1.34689859293$.  The Bohm$(3,3)_3$ metric
on $S^4\times S^3$ has $b_0\approx 0.14291337$ and
$t_c\approx 1.4691901856$.  The Bohm$(4,4)_2$ metric on $S^9$ has 
$b_0\approx 0.2851829$ and $t_c\approx 1.376730624$, whilst the
Bohm$(4,4)_3$ metric on $S^5\times S^4$ has $b_0\approx 0.09135$ and
$t_c\approx 1.5099148$.

   We can also treat the analysis of the non-compact Bohm metrics
described in section \ref{noncompsec} in a similar fashion.  Since
these are Ricci-flat solutions of the Einstein equations, the terms
involving the cosmological constant will be absent in (\ref{einstein})
and (\ref{constraint}), but otherwise all the formulae are analogous.
The short-distance Taylor expansions (\ref{taylor}) now become
\bea
a &=& t -\fft{q\, (q-1)}{6b_0^2\, p\, (p+1)}\, t^3
  + O(t^5)\,,\nn\\
b &=& b_0 + \fft{q-1}{2b_0\, (p+1)}\, t^2 + O(t^4)\,.
\label{taylor2}
\eea
Using this, taken to order $t^9$, to set initial data just outside the 
$S^q$ bolt at $t=0$, we again performed numerical integrations.  The
plots for the functions $a$ and $b$ in the representative example
$p=q=2$ are given at the end of this section.



\begin{figure}[hbp]
\centerline{\epsfxsize=3.4truein
\epsffile{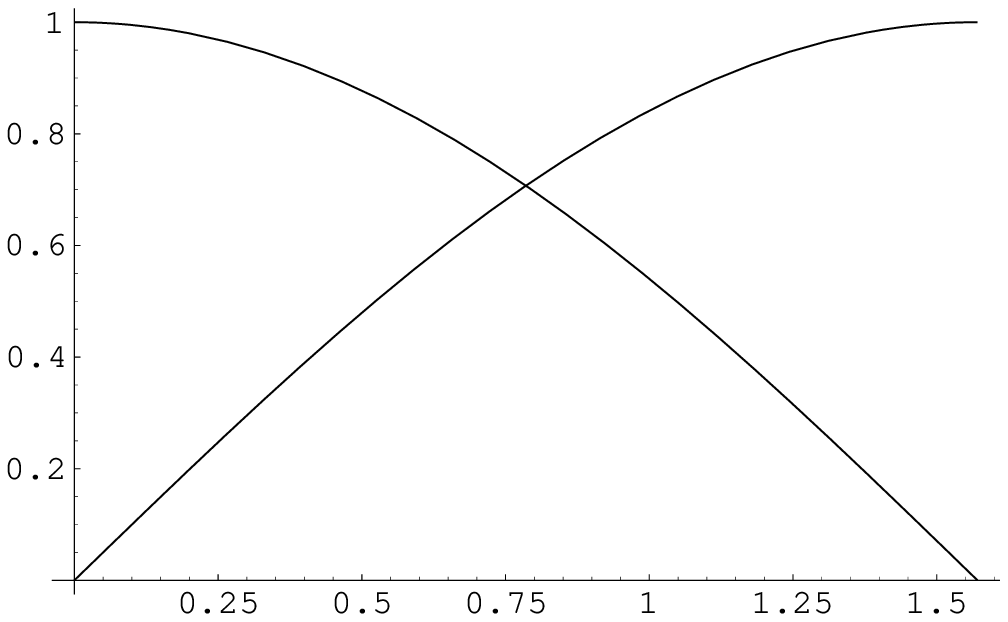}
\hspace{0.25in}
\epsfxsize=2.4truein
\epsffile{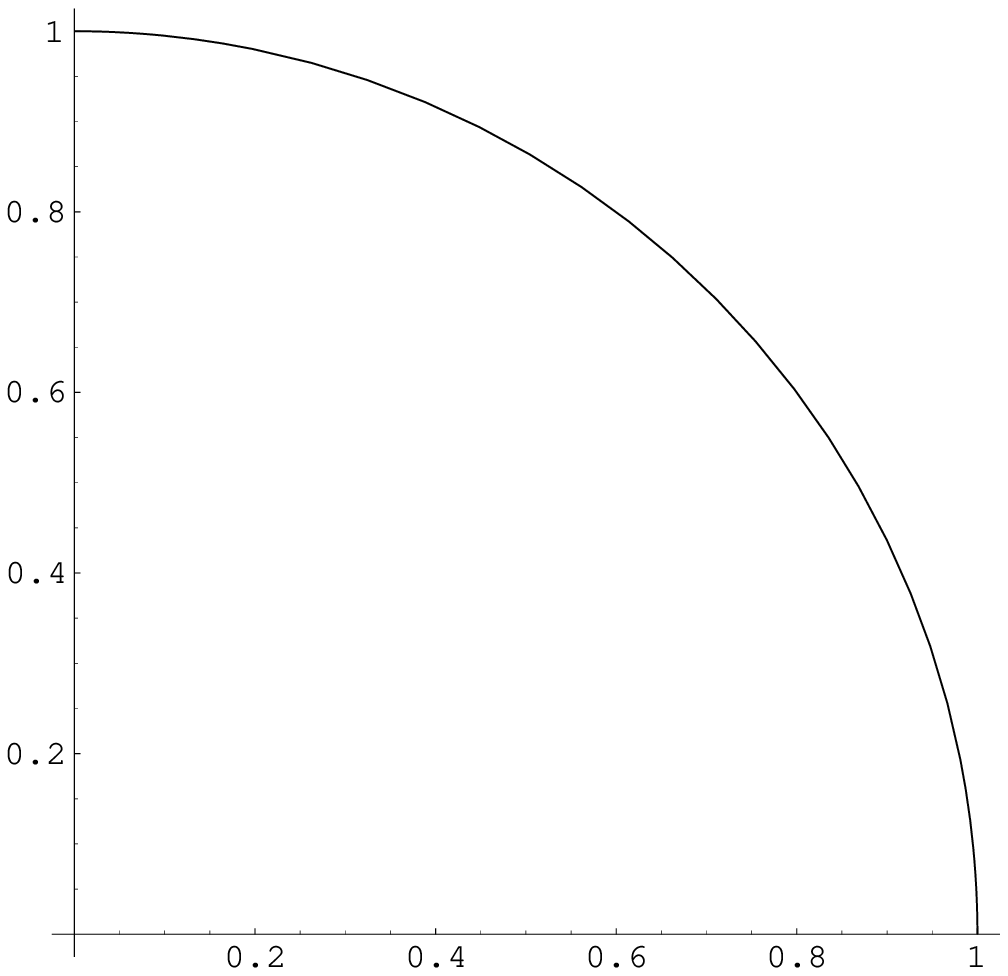}
}
\caption{The Bohm$(2,2)_0$ (standard) Einstein metric on $S^5$.  The
left-hand figure shows the metric coefficients $a$ and $b$ as
functions of the radial variable $t$.  The function $a$ vanishes at
$t=0$, and $b=b_0=1$ there.  The crossover occurs at $t=t_c=\ft14\pi$.
The right-hand figure is a parametric plot of $b$ vs. $a$. }
%
\vskip 1truein
\centerline{\epsfxsize=3.4truein
\epsffile{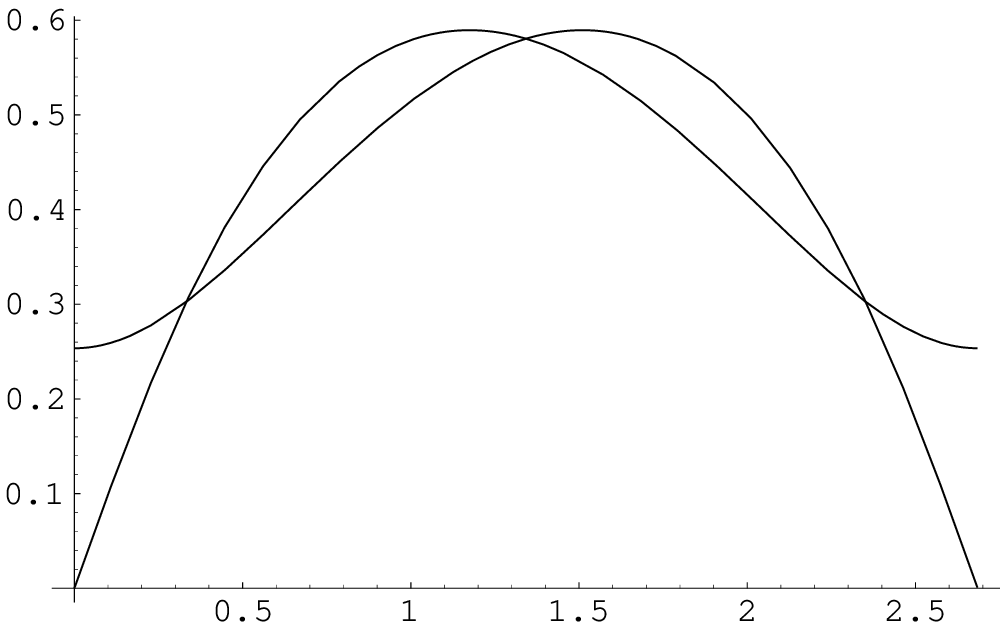}
\hspace{0.25in}
\epsfxsize=2.4truein
\epsffile{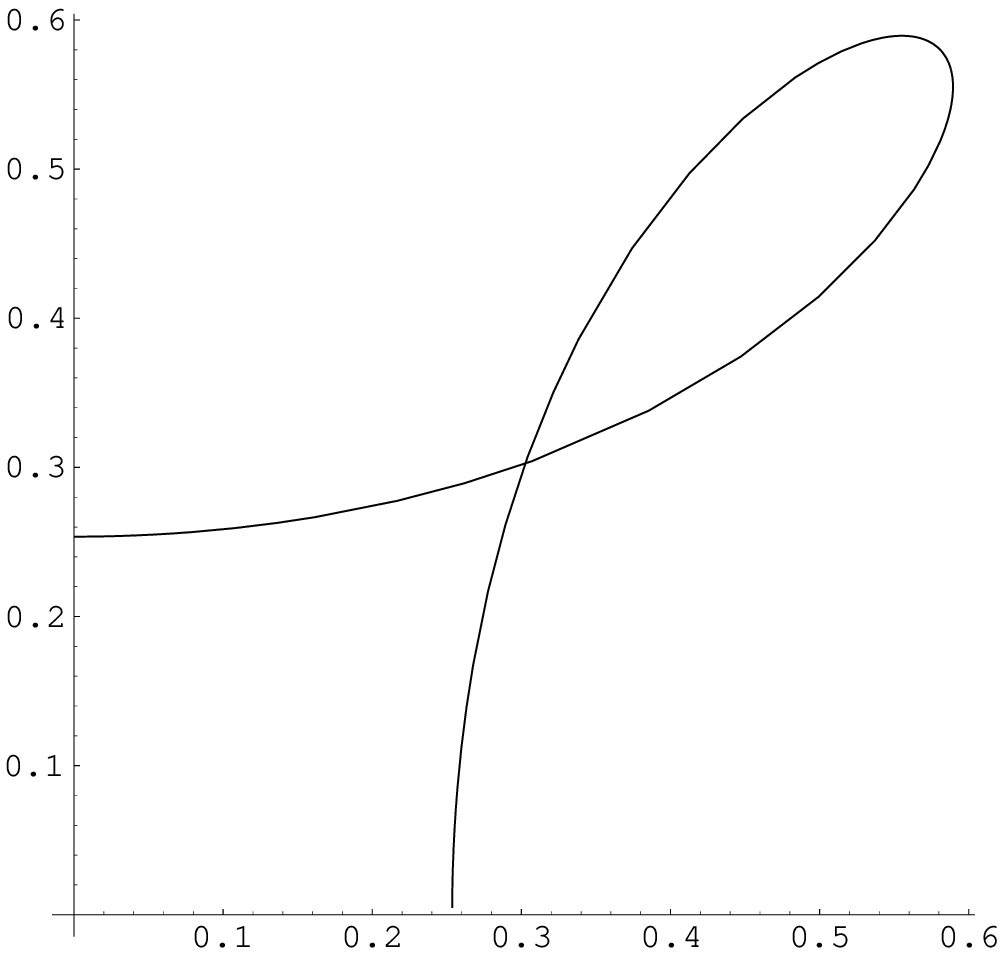}
}
\caption{The Bohm$(2,2)_2$ Einstein metric on $S^5$.  The left-hand figure
shows the metric coefficients $a$ and $b$ as functions of the radial
variable $t$.  At $t=0$ the function $a$ vanishes, and $b=b_0\approx
0.253554255$.  The mid-point is at $t_c\approx 1.34235319$.  
The right-hand figure is a parametric plot of $b$ vs. $a$. }
\end{figure}

\begin{figure}[htbp]
\centerline{\epsfxsize=3.4truein
\epsffile{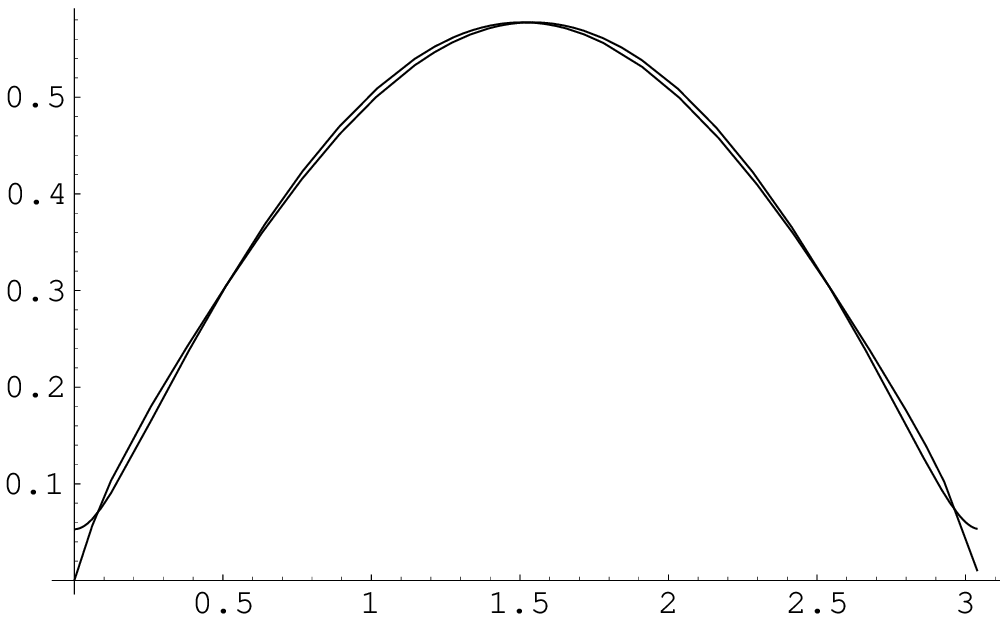}
\hspace{0.25in}
\epsfxsize=2.4truein
\epsffile{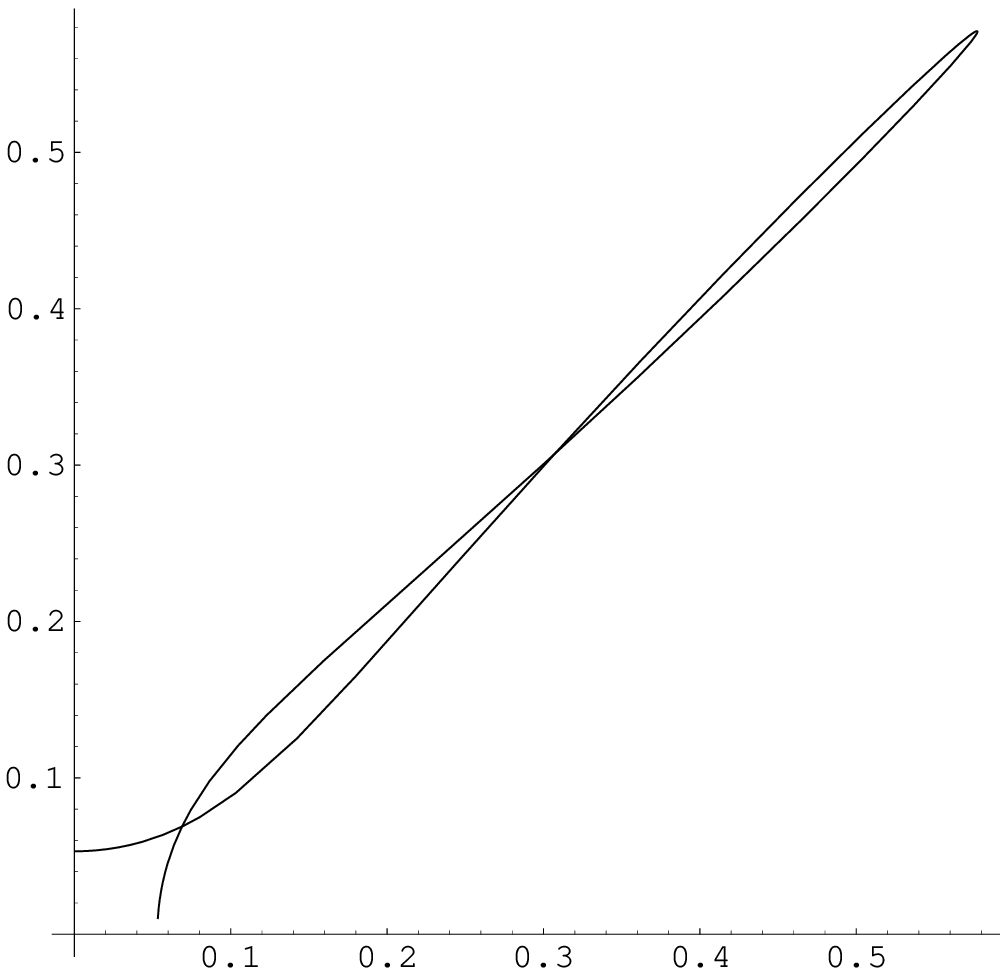}
}
\caption{The Bohm$(2,2)_4$ Einstein metric on $S^5$.  The function $b$ starts
at $b_0\approx 0.053054$, and the mid-point is at $t_c\approx 1.524951$.}
\end{figure}

\vskip 1truein

\begin{figure}[htbp]
\centerline{\epsfxsize=3.4truein
\epsffile{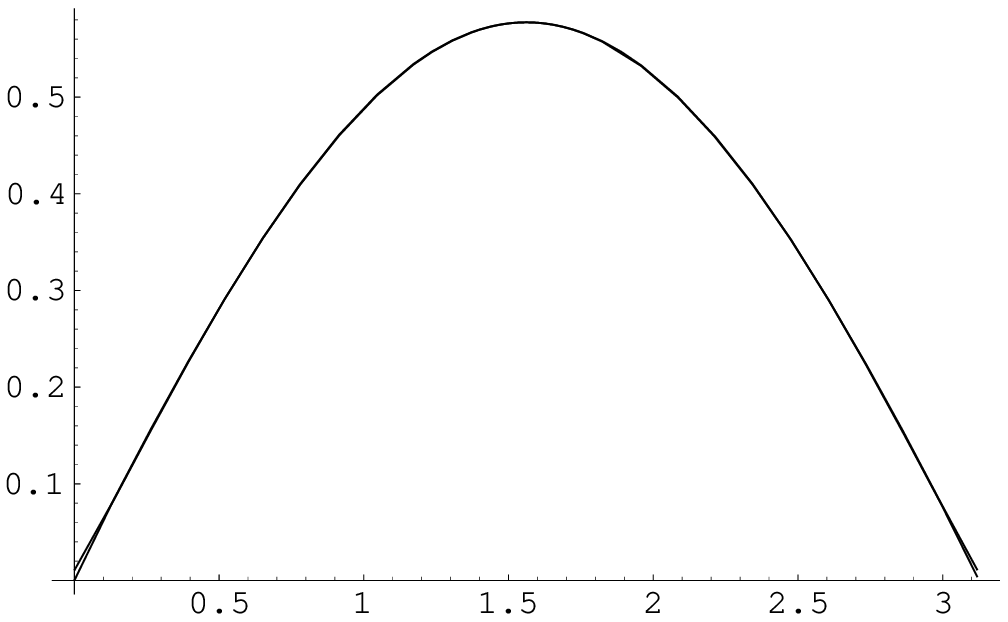}
\hspace{0.25in}
\epsfxsize=2.4truein
\epsffile{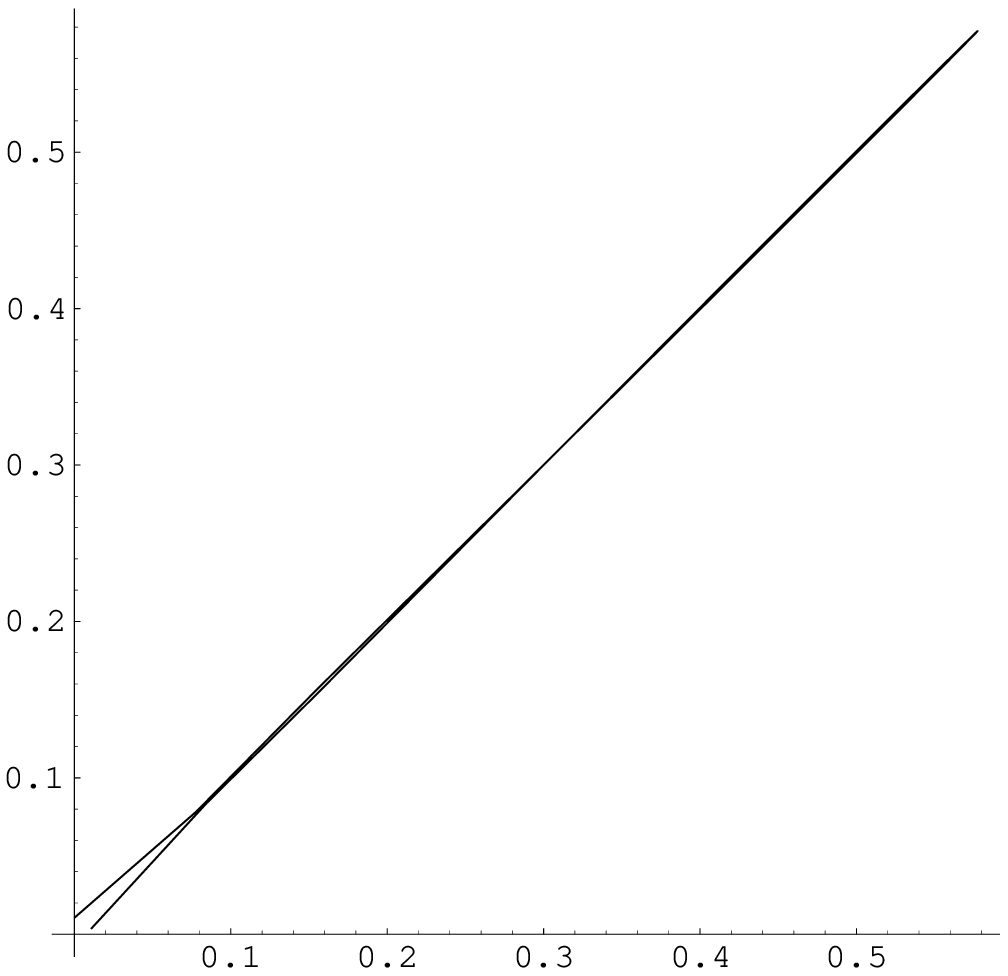}
}
\caption{The Bohm$_6$ Einstein metric on $S^5$.  The function $b$ starts 
at $b_0\approx 0.010503$, and the mid-point is at $t_c\approx 1.56174$.
}
\end{figure}

\clearpage


\begin{figure}[htbp]
\centerline{\epsfxsize=3.4truein
\epsffile{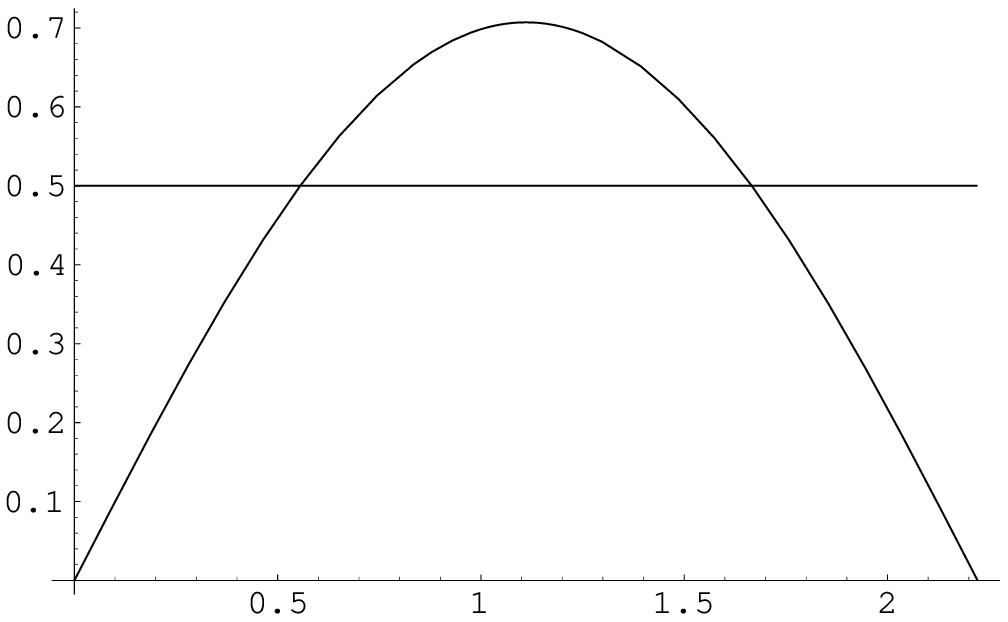}
\hspace{0.25in}
\epsfxsize=2.4truein
\epsffile{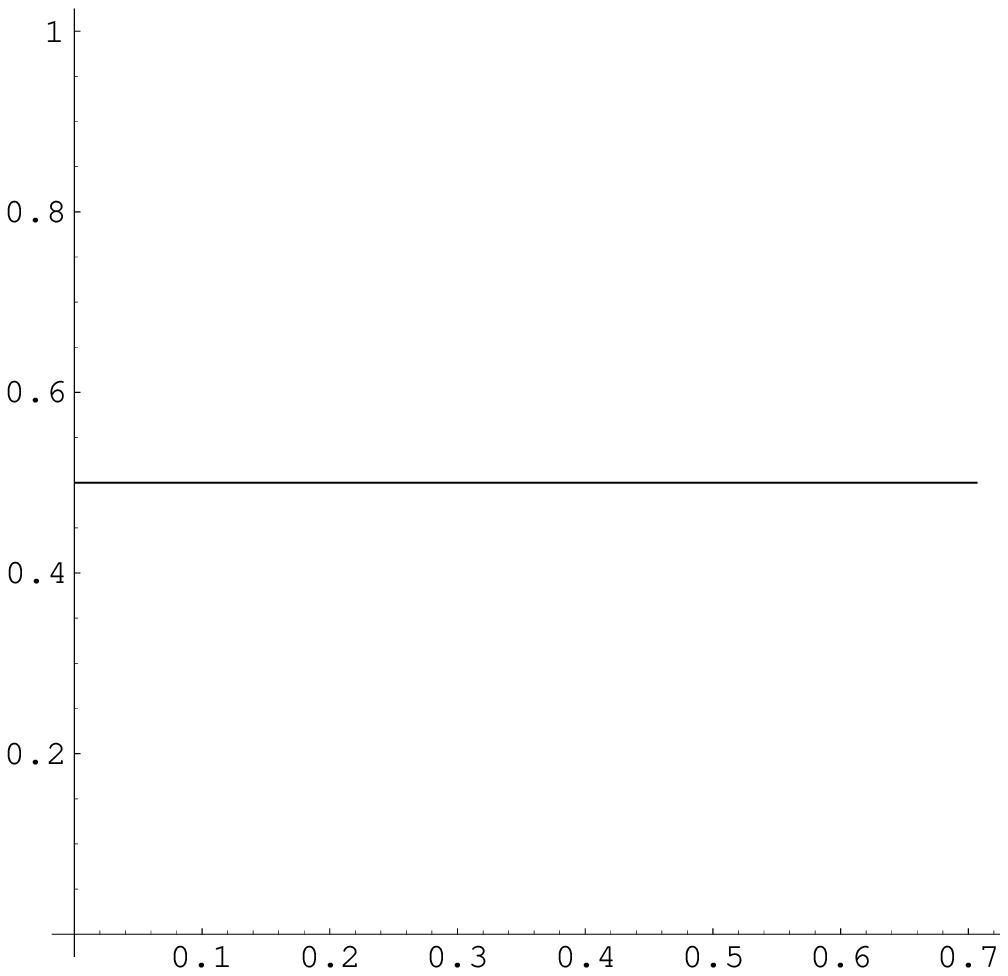}
}
\caption{The Bohm$(2,2)_1$ (standard) Einstein metric on $S^3\times S^2$.  
The left-hand figure
shows the metric coefficients $a$ and $b$ as functions of the radial
variable $t$.  At $t=0$ the function $a$ vanishes, and $b_0=\ft12$.  
The mid-point is at $t_0=\ft1{2\sqrt2}\, \pi$.  The right-hand
figure is a parametric plot of $b$ vs. $a$. }
\vskip 1truein
\centerline{\epsfxsize=3.4truein
\epsffile{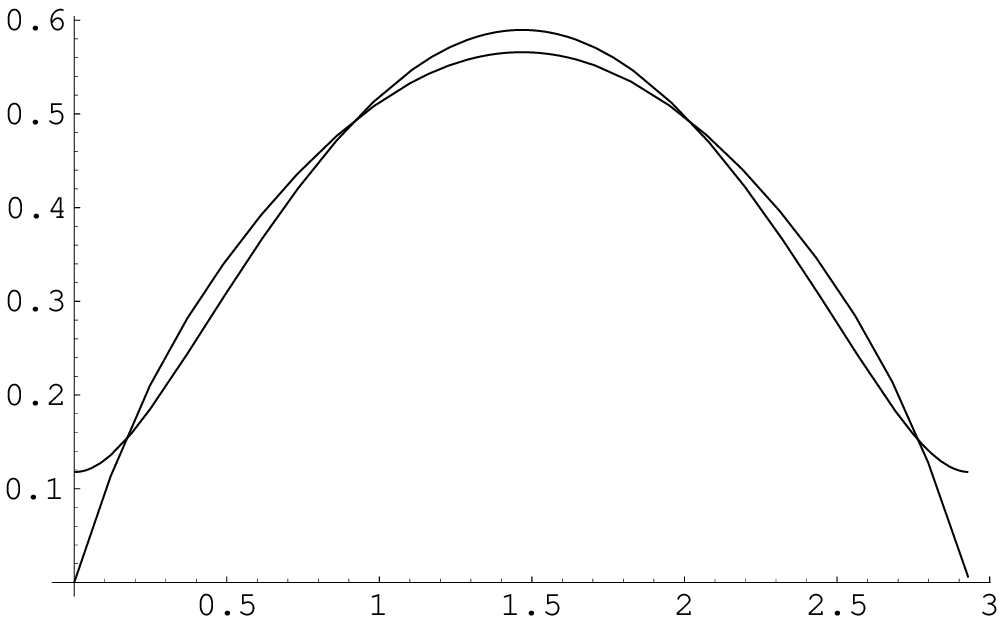}
\hspace{0.25in}
\epsfxsize=2.4truein
\epsffile{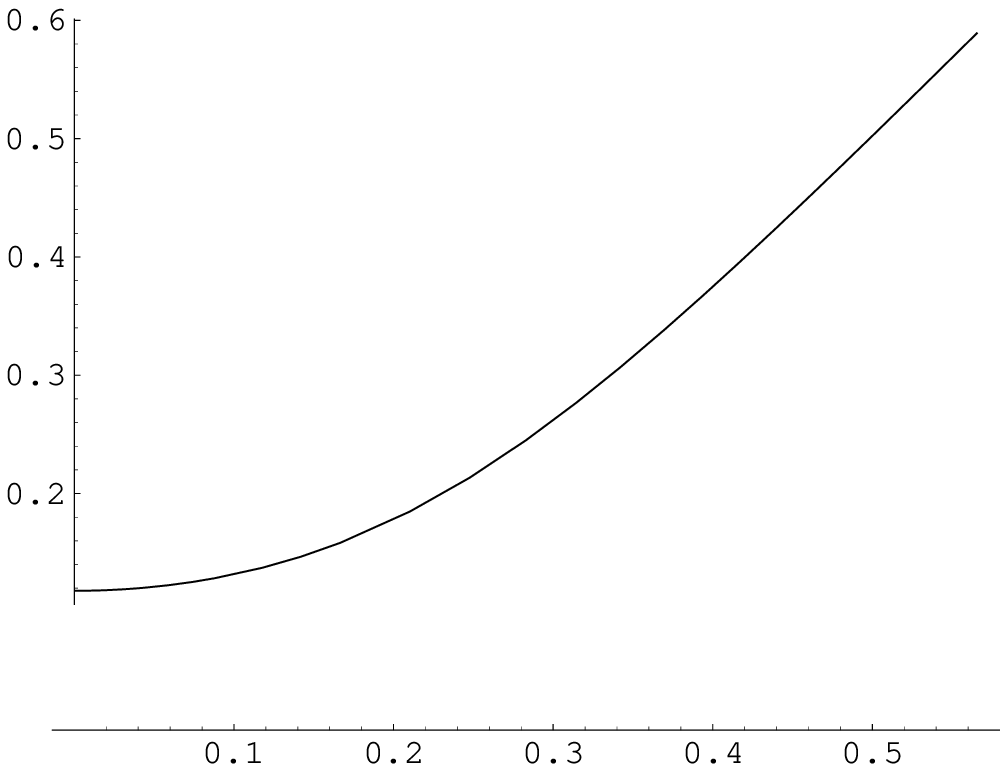}
}
\caption{The Bohm$(2,2)_3$ Einstein metric on $S^3\times S^2$.  
The left-hand figure
shows the metric coefficients $a$ and $b$ as functions of the radial
variable $t$.  At $t=0$ the function $a$ vanishes, and $b_0\approx
0.117794$. The mid-point is at $t_c\approx 1.46768843$.  The right-hand
figure is a parametric plot of $b$ vs. $a$. }
\end{figure}

\begin{figure}[htbp]
\centerline{\epsfxsize=3.4truein
\epsffile{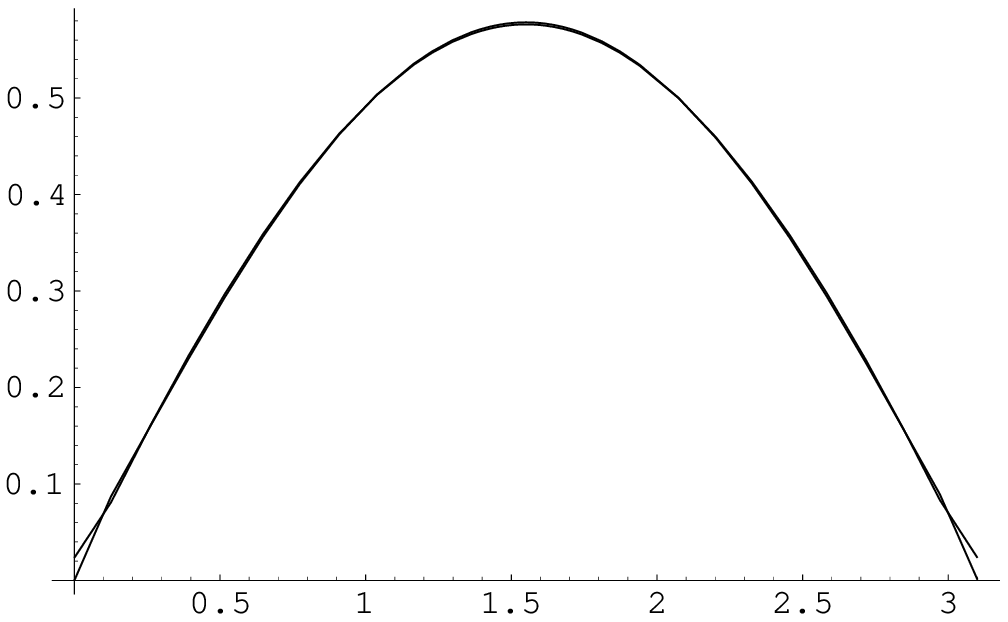}
\hspace{0.25in}
\epsfxsize=2.4truein
\epsffile{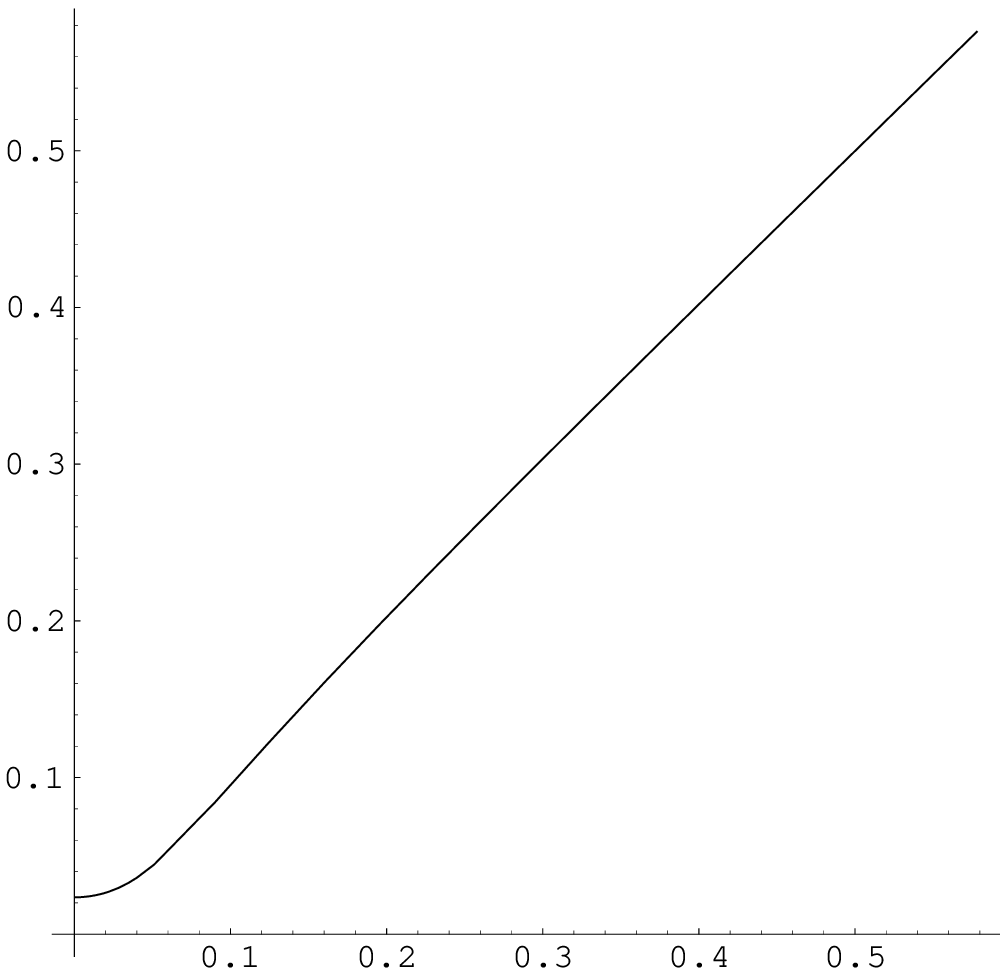}
}
\caption{The Bohm$(2,2)_5$ Einstein metric on $S^3\times S^2$.  
The left-hand figure
shows the metric coefficients $a$ and $b$ as functions of the radial
variable $t$.  At $t=0$ the function $a$ vanishes, and $b_0\approx
0.023571$. The mid-point is at $t_c\approx 1.550472593$.  The right-hand
figure is a parametric plot of $b$ vs. $a$. }
\end{figure}


\begin{figure}[htbp]
\centerline{\epsfxsize=3.4truein
\epsffile{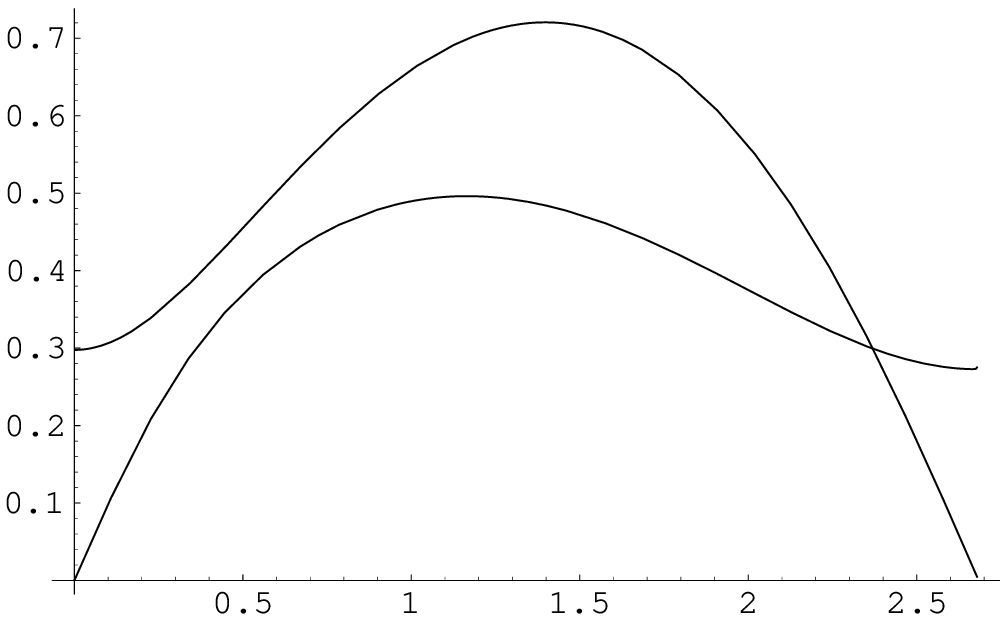}
\hspace{0.25in}
\epsfxsize=2.4truein
\epsffile{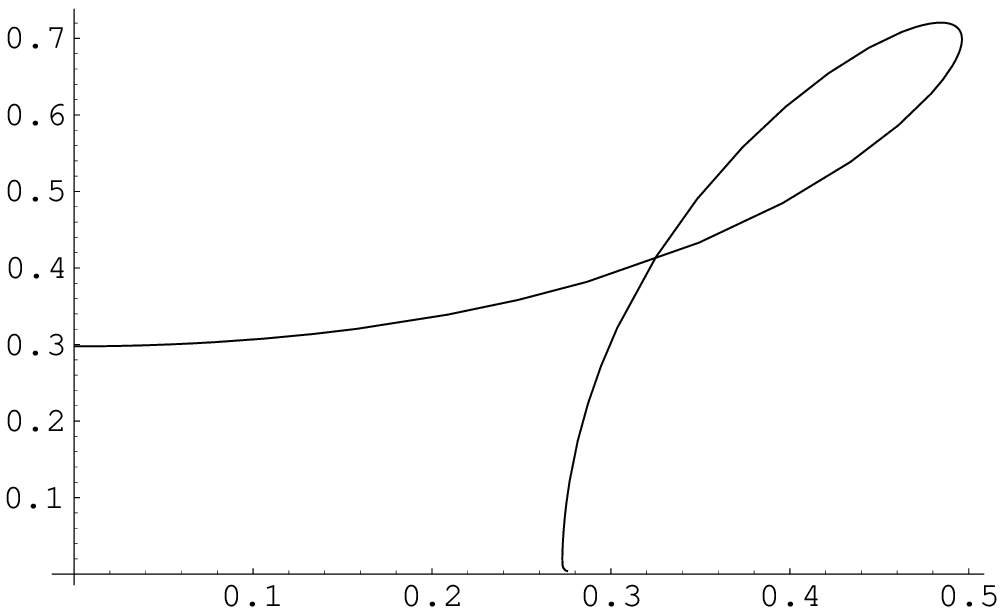}
}
\caption{The Bohm$(2,3)_2$ Einstein metric on $S^6$.  
The left-hand figure
shows the metric coefficients $a$ and $b$ as functions of the radial
variable $t$.  At $t=0$ the function $a$ vanishes, and $b_0\approx
0.297647$. The endpoint is at $t_f\approx 2.68296$.  The right-hand
figure is a parametric plot of $b$ vs. $a$. }
\end{figure}


\begin{figure}[htbp]
\centerline{\epsfxsize=3.4truein
\epsffile{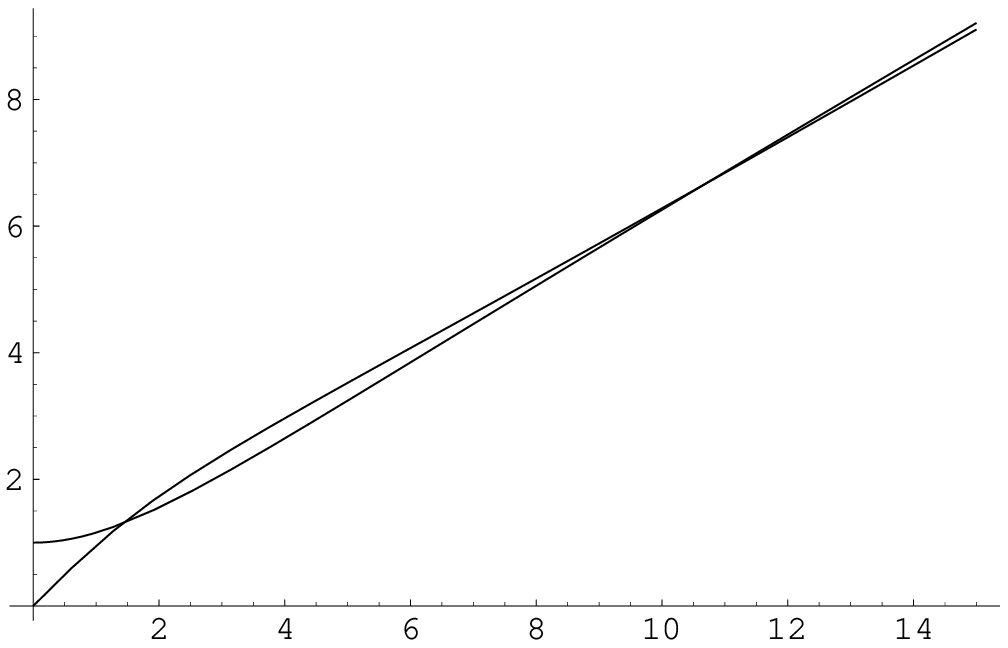}
\hspace{0.25in}
\epsfxsize=2.4truein
\epsffile{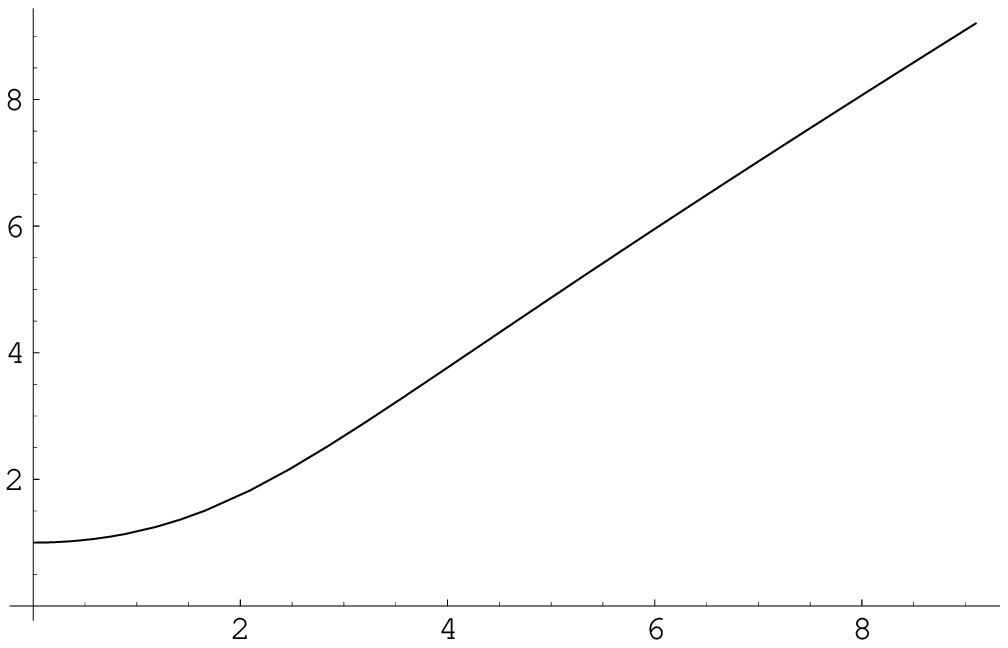}
}
\caption{The non-compact Ricci-flat Bohm metric on $\R^3\times S^2$.  
The left-hand figure
shows the metric coefficients $a$ and $b$ as functions of the radial
variable $t$.  At $t=0$ the function $a$ vanishes, and $b_0$ is 
taken to be 1. The right-hand
figure is a parametric plot of $b$ vs. $a$. }
\end{figure}


\clearpage

\begin{thebibliography}{99}

\bibitem{Giddings} S.B. Giddings and S. Thomas, {\it High energy
colliders as black holes factories: the end of short distance
physics}, Phys. Rev. {\bf D65} (2002) 056010, {\tt hep-th/0106219}.

\bibitem{Dimopoulos} S. Dimopoulos and G. Landsberg, {\it Black holes at
the LHC}, Phys. Rev. Lett. {\bf 87} (2201) 161602, {\tt
hep-th/0106219}.

\bibitem{GibbonsHawking1} G.W. Gibbons and S.W. Hawking, 
{\it Cosmological event horizons,
thermodynamics and particle creation}, Phys Rev {\bf D15} (1977) 2738-2751.

\bibitem{Ida1} G.W. Gibbons, D. Ida and T. Shiromizu, {\it Uniqueness and
nonuniqueness of static black holes in higher dimensions},
Phys. Rev. Lett. {\bf 89} (2002) 041101. {\tt hep-th/0206049}

\bibitem{Ida2} G.W. Gibbons, D. Ida and T. Shiromizu, {\it Uniqueness of
(dilatonic) charged black holes and black $p$-branes in higher
dimensions}, {\tt hep-th/0206136}.

\bibitem{bohm} C. Bohm, {\it Inhomogeneous Einstein metrics on
low-dimensional spheres and other low-dimensional spaces},
Invent. Math. {\bf 134} (1998) 145.

\bibitem{gh} G.W. Gibbons and S.A. Hartnoll, {\it A gravitational instability
in higher dimensions}, {\tt hep-th/0206202}.

\bibitem{BoucherGibbons} W. Boucher and G.W. Gibbons, {\it Cosmic
baldness}
in {\it The Very Early Universe}, eds. G.W. Gibbons, S.W. Hawking and
S.T.C. Siklos, Cambridge University Press (1983).

\bm{pagepope}D.N. Page and C.N. Pope,
{\it Stability analysis of compactifications of $D = 11$ supergravity 
with $SU(3) X SU(2) X U(1)$ symmetry},
Phys. Lett. {\bf B145} (1984) 337.

\bm{pagepope2} D.N. Page and C.N. Pope,
{\it Which compactifications of $D = 11$ supergravity are stable?},
Phys. Lett. {\bf B144} (1984) 346.

\bibitem{Kol} B. Kol, {\it Topology change in general relativity and
the black-hole black-string transition}, {\tt hep-th/0206220}.

\bibitem{dnp} M.J. Duff, B.E.W. Nilsson and C.N. Pope, {\it The criterion
for vacuum stability in Kaluza-Klein supergravity}, Phys. Lett. {\bf B139}
(1984) 154.

\bibitem{dfghm} O. DeWolfe, D.Z. Freedman, S.S. Gubser, G.T. Horowitz
and I. Mitra, {\it Stability of $AdS_p\times M_q$ Compactifications
without supersymmetry},
Phys. Rev. {\bf D65} (2002) 064033, {\tt hep-th/0105047}.

\bibitem{wiltshire} D.L. Wiltshire. {\it Spherically symmetric
solutions of Einstein-Maxwell theory with a Gauss-Bonnet term},
Phys. Lett. {\bf B169} (1986) 36.

\bibitem{bf} P. Breitenlohner and D.Z. Freedman, {\it Positive energy
in anti-de Sitter backgrounds and gauged extended supergravity},
Phys. Lett. {\bf B139} (1984) 154.

\bibitem{mt} L. Mezincescu and P.K. Townsend, {\it Stability at a
local maximum in higher dimensional anti-de Sitter space and
applications to supergravity}, Ann. Phys. {\bf 160} (1985) 406.

\bibitem{Bohm2} C. Bohm, {\it Non-compact cohomogeneity
one Einstein Manifolds}, Bull. Soc. Math. France, {\bf 127} (1999) 135-177.

\bibitem{Galicki1} C.P. Boyer and K.Galicki, {\it New Einstein metrics
in dimension five}, J. Diff. Geom. {\bf 57} (2001) 443-463 {\tt DG/0003174}.

\bibitem{Galicki2} C.P. Boyer and K.Galicki, {\it On Sasakian-Einstein
geometry}, Int. J. Math {\bf 11} (2000) 873-909, {\tt DG/9811098}

\bibitem{Galicki3} C.P. Boyer, K. Galicki and M. Nakamaye, {\it
Sasakian-Einstein structures on $9\#(S^2\times S^3)$}, 
{\tt DG/0102181}.

\bibitem{Galicki4} C.P. Boyer, K. Galicki and M.Nakamaye, {\it On the
geometry of Sasakian-Einstein 5-Manifolds}, {\tt DG/0012047}.

\bibitem{Herzog} A. Bergman and C. Herzog, {\it The volume of some
non-spherical horizons and the AdS/CFT correspondence},  JHEP 0201(2002)
030  {\tt hep-th/0108020}.

\bibitem{GibbonsB} G.W. Gibbons, {\it Tunnelling with a negative
cosmological constant}, Nucl. Phys. {\bf B472} (1996) 683-708 {\tt
hep-h/9601075}.

\bibitem{GibbonsHartle} G.W. Gibbons and J.B. Hartle, {\it Real
tunnelling geometries and the large-scale topology of the universe},
Phys Rev {\bf D42} (1990) 2458-2568.

\bibitem{Bishop} R.L. Bishop, {\it A relation between volume, mean
curvature and diameter}, Notices Amer. Math. Soc. {\bf 10} (1963) 364.

\bibitem{GibbonsC} G.W. Gibbons, {\it Real tunnelling geometries},
Class. Quant. Grav. {\bf 15} (1998) 2605-2612.

\bibitem{Gursky} M.J. Gursky, {\it Four-manifolds with $\delta W^+=0$
and Einstein constants of the sphere}, Math. Ann. {\bf 318} (2000)
417-431.

\bibitem{GibbonsRuback} G.W. Gibbons and P.J. Ruback, {\it Classical
gravitons and their stability in higher dimensions}, Phys. Lett. {\bf
B171} (1986) 390.

\bibitem{Singer} M.A. Singer, {\it Positive Einstein metrics with small
$L^{n \over 2}$-norm of the Weyl tensor}, Diff. Geom. Appl. {\bf 2}
(1992) 269-274.

\bibitem{Boucher} W. Boucher, in {\it Classical General Relativity},
eds. W. Bonnor, J. Islam and M.A.H. MacCallum, Cambridge University
Press (1983).

\bibitem{Friedrichs} H. Friedrich, {\it Existence and structure of
past asymptotically simple solutions of Einstein's field equations with
positive cosmological constant}, J Geom Phys {\bf 3} (1986) 101-117.

\bibitem{Prestidge} T. Prestige, {\it Dynamic and thermodynamic
stability and negative modes in Schwarzschild-anti-de Sitter},
Phys. Rev. {\bf D61} (2000) 084002, {\tt hep-th/9907163}.

\bibitem{GregoryRoss} J.P. Gregory and S.F. Ross, {\it Stability and
the negative mode for Schwarzschild in a finite cavity},
Phys. Rev. {\bf D64} (2001) 124006, {\tt hep-th/0106220}.

\bibitem{Reall} H.S. Reall, {\it Classical and thermodynamic stability
of black branes}, Phys. Rev. {\bf D64} (2001) 044005, {\tt hep-th/0104071}.

\bibitem{GubserMitra1} S.S. Gubser and I. Mitra, {\it Instability of
charged black holes in anti-de Sitter space}, {\tt hep-th/0009126}.

\bibitem{GubserMitra2} S.S. Gubser and I. Mitra, {\it The evolution of
unstable black holes in anti-de Sitter space}, JHEP 0108 (2001) 018,
{\tt hep-th/0011127}.

\bibitem{HawkingPage} S.W. Hawking and D.N. Page, {\it Thermodynamics of
black holes in anti-de Sitter space}, Commun. Math. Phys. {\bf 87}
(1983) 577.

\bibitem{GibbonsHawking2} G.W. Gibbons, {\it Classification of
gravitational instanton symmetries}, Comm. Math. Phys. {\bf 66} (1979)
291-301.

\bibitem{hartnoll} S.A. Hartnoll, {\it Axisymmetric non-abelian BPS
monopoles from $G_2$ metrics}, Nucl. Phys. {\bf B631} (2002) 325-341,
{\tt hep-th/0112235}.

\end{thebibliography}
\end{document}